\documentclass{article}
\usepackage{makeidx}
\usepackage{amsfonts}
\usepackage{amsmath}

\newtheorem{theorem}{Theorem}

\newtheorem{corollary}[theorem]{Corollary}

\newtheorem{example}[theorem]{Example}

\newtheorem{lemma}[theorem]{Lemma}

\newtheorem{proposition}[theorem]{Proposition}
\newtheorem{remark}[theorem]{Remark}

\newenvironment{proof}[1][Proof]{\noindent\textbf{#1.} }{\ \rule{0.5em}{0.5em}}
\input{tcilatex}

\begin{document}

\title{A Weyl Calculus on Symplectic Phase Space}
\author{Maurice A de Gosson \\
%EndAName
Universit\"{a}t Potsdam, \\
Inst. f. Mathematik, \\
Am Neuen Palais 10, D-14415 Potsdam \\
and\\
Universidade de S\~{a}o Paulo\\
Departamento de Matem\~{a}tica, \\
CEP 05508-900 \ S\~{a}o Paulo}
\maketitle

\begin{abstract}
We define and study a metaplectically covariant class of pseudo-differential
operators acting on functions on symplectic space and generalizing a
modified form of the usual Weyl calculus. This construction requires a
precise calculation of the twisted Weyl symbol of a class of generators of
the metaplectic group and the use of a Conley--Zehnder type index for
symplectic paths, defined without restrictions on the endpoint. Our calculus
is related the usual Weyl calculus using a family of isometries of $L^{2}(%
\mathbb{R}^{n})$ on closed subspaces of $L^{2}(\mathbb{R}^{2n})$ and to an
irreducible representation of the Heisenberg algebra distinct from the usual
Schr\"{o}dinger representation.
\end{abstract}

\tableofcontents

\textit{AMS\ Classification} (2000): 81S30, 43A65, 43A32


\noindent \textbf{Acknowledgements.} This work has been supported by the
grant 2005/51766-7 of the Funda\c{c}\~{a}o de Amparo \`{a} Pesquisa do
Estado de S\~{a}o Paulo (FAPESP). I thank P. Piccione (S\~{a}o Paulo) for
his kind and congenial hospitality.

\section{Introduction}

It is part of the mathematical folklore to describe the metaplectic
representation of the real symplectic group $\limfunc{Sp}(Z,\sigma )$ ($Z=%
\mathbb{R}^{2n}$, $\sigma $ the standard symplectic form) in terms of
unitary operators acting on functions in $n$ variables; these variables are
either the \textquotedblleft position coordinates\textquotedblright\ $%
x=(x_{1},...,x_{n})$ or the dual \textquotedblleft momentum
coordinates\textquotedblright\ $p=(p_{1},...,p_{n})$, or a mixture of both
containing no \textquotedblleft conjugate pairs\textquotedblright\ $x_{j}$, $%
p_{j}$. There is thus a discrepancy between symplectic \textit{geometry},
where $\limfunc{Sp}(Z,\sigma )$ acts on phase-space points depending on $2n$
variables $(x,p)$, and symplectic \textit{harmonic analysis} where the
metaplectic group $\limfunc{Mp}(Z,\sigma )$ acts on functions of half as
many variables. This state of affairs is rarely questioned by quantum
physicists: the metaplectic representation intervening both in an
\textquotedblleft active\textquotedblright\ and a \textquotedblleft
passive\textquotedblright\ way in quantum mechanics, it is comforting for
them that $\limfunc{Mp}(Z,\sigma )$ can only be seen, to paraphrase Dirac,
\textquotedblleft with the $x$-eye or the $p$-eye\textquotedblright : for
them the uncertainty principle prohibits the existence of a
quantum-mechanical phase space.

It turns out that it is perfectly possible to construct a metaplectic
representation $\limfunc{Mp}_{\text{ph}}(Z,\sigma )$ of $\limfunc{Sp}%
(Z,\sigma )$ acting on functions depending on the phase space variable $%
z=(x,p)$; to this representation is associated a pseudo-differential
calculus on $Z$ which is symplectically covariant under conjugation with
elements of $\limfunc{Mp}_{\text{ph}}(Z,\sigma )$. There are actually at
least two options for doing this. There is the easy way, which consists in
constructing an isometry $U$ of $L^{2}(\mathbb{R}^{n})$ on a subspace of $%
L^{2}(\mathbb{R}^{2n})$ (for instance the \textquotedblleft coherent state
representation\textquotedblright , familiar to physicists), and to make $%
\widehat{S}\in \limfunc{Mp}(Z,\sigma )$ act on $L^{2}(Z)$ by intertwining it
with $U$. This straightforward approach has the disadvantage that it is
tautological: we do not obtain a true action of $\limfunc{Mp}(Z,\sigma )$ on
all of $L^{2}(Z)$, but only on a subspace isometric to $L^{2}(\mathbb{R}%
^{n}) $; it is certainly not obvious what sense to give to $\widehat{S}f$
for arbitrary $f\in L^{2}(Z)$. We will follow another way, which requires
some more work, but which is in the end far more rewarding. It consists in
two steps: one first writes the elements of a set of generators of $\limfunc{%
Mp}(Z,\sigma )$ in Weyl form%
\begin{equation*}
\widehat{S}=\left( \tfrac{1}{2\pi }\right) ^{n}\int a_{\sigma
}(z_{0})e^{-i\sigma (\widehat{z},z_{0})}dz_{0}
\end{equation*}%
where $a_{\sigma }$ is the twisted symbol of $\widehat{S}$ (symplectic
Fourier transform of the usual symbol) and $\widehat{z}=(x,-i\partial _{x})$%
. One then observes that the action of $e^{-i\sigma (\widehat{z},z_{0})}f$
is, for $f\in \mathcal{S}(\mathbb{R}^{n})$, the time-one solution to Schr%
\"{o}dinger's equation%
\begin{equation*}
i\partial _{t}\psi =\sigma (\widehat{z},z_{0})\psi \text{ \ , \ }\psi
(x,0)f(x)
\end{equation*}%
and is hence explicitly given by the formula%
\begin{equation*}
e^{-i\sigma (\widehat{z},z_{0})}f(x)=e^{i(\left\langle p_{0},x\right\rangle -%
\tfrac{1}{2}\left\langle p_{0},x_{0}\right\rangle )}f(x-x_{0})\text{;}
\end{equation*}%
this can be rewritten as%
\begin{equation*}
e^{-i\sigma (\widehat{z},z_{0})}f(x)=\widehat{T}(z_{0})f(x)
\end{equation*}%
where 
\begin{equation*}
\widehat{T}(z_{0})=e^{i(\left\langle p_{0},x\right\rangle -\tfrac{1}{2}%
\left\langle p_{0},x_{0}\right\rangle )}T(z_{0})
\end{equation*}%
is the Heisenberg--Weyl operator familiar from the theory of the Heisenberg
group (here $T(z_{0})f(x)=f(x-x_{0})$). One next makes the (very pedestrian)
observation that at this point there is no need to limit the range of the
operators $\widehat{T}(z_{0})$ to functions of $x$, so one extends them by
defining, for $F\in \mathcal{S}(Z)$, 
\begin{equation}
\widehat{T}(z_{0})F(z)=e^{i(\left\langle p_{0},x\right\rangle -\tfrac{1}{2}%
\left\langle p_{0},x_{0}\right\rangle )}F(z-z_{0})\text{.}  \label{obvious}
\end{equation}

The procedure just outlined was actually hinted at in the first part of the
seminal paper by Grossmann \textit{et al}. \cite{Grossmann}, but not fully
exploited; in this paper we will actually use a slight variant of the
construction above: instead of defining the phase-space operators by bluntly
extending the domain of $\widehat{T}(z_{0})=e^{-i\sigma(\widehat{z},z_{0})}$%
, we will use the operators $\widehat{T}_{\text{ph}}(z_{0})$ defined by%
\begin{equation*}
\widehat{T}_{\text{ph}}(z_{0})F(z)=e^{-\tfrac{i}{2}\sigma(z,z_{0})}F(z-z_{0})%
\text{;}
\end{equation*}
equivalently%
\begin{equation*}
\widehat{T}_{\text{ph}}(z_{0})=e^{-i\sigma(\widehat{z}_{\text{ph}},z_{0})}
\end{equation*}
where $\widehat{z}_{\text{ph}}$ is the operator on $\mathcal{S}(Z)$ defined
by 
\begin{equation}
\widehat{z}_{\text{ph}}=(\tfrac{1}{2}x+i\partial_{p},\tfrac{1}{2}%
p-i\partial_{x})\text{.}  \label{quru}
\end{equation}
Notice that these modified Heisenberg--Weyl operators $\widehat{T}_{\text{ph}%
}(z_{0})$ satisfy the same commutation and product relations 
\begin{align}
\widehat{T}_{\text{ph}}(z_{0})\widehat{T}_{\text{ph}}(z_{1}) &
=e^{-i\sigma(z_{0},z_{1})}\widehat{T}_{\text{ph}}(z_{1})\widehat{T}_{\text{ph%
}}(z_{0})  \label{noco1} \\
\widehat{T}_{\text{ph}}(z_{0}+z_{1}) & =e^{-\frac{i}{2}\sigma(z_{0},z_{1})}%
\widehat{T}_{\text{ph}}(z_{0})\widehat{T}_{\text{ph}}(z_{1})  \label{noco2}
\end{align}
as the operators $\widehat{T}(z_{0})$ and will therefore allow the
construction of an irreducible unitary representation of the Heisenberg
group $\mathbf{H}_{n}$. This procedure allows us to associate to an
arbitrary Weyl operator%
\begin{equation*}
\widehat{A}=\left( \tfrac{1}{2\pi}\right) ^{n}\int
a_{\sigma}(z_{0})e^{-i\sigma(\widehat{z},z_{0})}dz_{0}
\end{equation*}
the \textquotedblleft phase-space operator\textquotedblright\ 
\begin{equation*}
\widehat{A}_{\text{ph}}=\left( \tfrac{1}{2\pi}\right) ^{n}\int a_{\sigma
}(z_{0})e^{-i\sigma(\widehat{z}_{\text{ph}},z_{0})}dz_{0}\text{;}
\end{equation*}
the operators $\widehat{A}_{\text{ph}}$ and $\widehat{A}$ are coupled by the
formula%
\begin{equation}
\widehat{A}_{\text{ph}}W^{\prime}(f,\overline{g})=W^{\prime}(\widehat {A}f,%
\overline{g})  \label{w}
\end{equation}
for all $f,\phi\in\mathcal{S}(\mathbb{R}^{n})$; here $W^{\prime}(f,\overline{%
g})(z)$ is a re-scaled variant of $W(f,\overline{g})$, the Wigner--Moyal
transform of the pair $(f,\overline{g})$ (Proposition \ref{propun}). An
essential feature of this correspondence is that the usual metaplectic
covariance of Weyl calculus is preserved: if we replace the symbol $a$ by $%
a\circ S$ where $S\in\limfunc{Sp}(Z,\sigma)$ then $\widehat {A}_{\text{ph}}$
is replaced by $\widehat{S}_{\text{ph}}^{-1}\widehat {A}_{\text{ph}}\widehat{%
S}_{\text{ph}}$.

This choice of definition of phase space operators, using $\widehat {T}_{%
\text{ph}}(z_{0})=e^{-i\sigma(\widehat{z}_{\text{ph}},z_{0})}$ instead of $%
\widehat{T}(z_{0})=e^{-i\sigma(\widehat{z},z_{0})}$, is not arbitrary, even
if it is not the only possible from a logical point of view. It has at least
two major advantages:

\begin{itemize}
\item The first advantage is that our choice makes the relationship between
the operators $\widehat{A}_{\text{ph}}$ with the Wigner--Moyal transform
very straightforward and allows the use of an already existing and
well-studied machinery. The more \textquotedblleft
obvious\textquotedblright\ definition using (\ref{obvious}) would instead
lead to technical complications; to be able to do reasonably easy
computations one would in the end anyway have to express the intertwining
formula in terms of Wigner--Moyal transform, at the cost of the appearance
of an unwanted exponential factor which would haunt us throughout the
calculation;

\item The second advantage, which is related to the first, is that it makes
the study of domains somewhat easier. As we will (briefly) discuss in the
Conclusion to this article one of the main applications of the theory we
sketch might well be quantum mechanics (Weyl calculus was after all designed
for this purpose). Assume that $\widehat{A}$ is, say, a unitary isometry of $%
\mathcal{S}(\mathbb{R}^{n})$ (it is the case if for instance $\widehat{A}\in%
\limfunc{Mp}(Z,\sigma)$). If we fix $g$ in the intertwining formula (\ref{w}%
) and let $f$ run through $\mathcal{S}(\mathbb{R}^{n})$ then $\widehat{A}_{%
\text{ph}}W^{\prime}(f,\overline{g})$ will describe a certain subspace of $%
\mathcal{S}(Z)$. Suppose in particular $g$ is a normalized Gaussian; then
that subspace consists of a very simple set of functions, namely those
functions $F$ such that $e^{\frac{1}{2}|z|^{2}}F$ is anti-analytic (Example %
\ref{theorange}).\bigskip
\end{itemize}

This article is structured as follows:

\begin{itemize}
\item In \S 2 we begin by recalling the definition and properties of the
Kashiwara--Wall signature of a triple of Lagrangian planes; we thereafter
review the theory of the Arnol'd--Leray--Maslov (ALM) index for pairs of
Lagrangian paths, and its by-product, the relative symplectic Maslov index
useful in the theory of the metaplectic group. We take the opportunity to
show on a few examples that these indices contain as particular cases some
other intersection indices appearing in the literature.

\item In \S 3 we define a new symplectic index, denoted by $\nu $, and
related to the familiar Conley--Zehnder index, but relaxed of any
non-degeneracy conditions on the endpoint of the path. The properties of a
\textquotedblleft symplectic Cayley transform\textquotedblright\ allow us to
relate that index $\nu $ to the relative Maslov index corresponding to a
particular polarization of the symplectic space. This property is
interesting \textit{per se }and could perhaps allow applications to the
theory of periodic Hamiltonian orbits; this possibility will however not be
investigated here in order to keep the length of the article within
reasonable limits;

\item In \S 4 we begin by reviewing the standard theory of the metaplectic
group $\limfunc{Mp}(Z,\sigma )$ and of its Maslov index. We then define a
family of unitary Weyl operators $\widehat{R}_{\nu }(S)$ parametrized by $%
S\in \limfunc{Sp}(Z,\sigma )$ such that $\det (S-I)\neq 0$ and $\nu \in 
\mathbb{R}$. These operators, which can be written in the very simple form\ 
\begin{equation*}
\widehat{R}_{\nu }(S)=\left( \tfrac{1}{2\pi }\right) ^{n}i^{\nu }\sqrt{|\det
(S-I)|}\int_{Z}\widehat{T}(Sz)\widehat{T}(-z)dz
\end{equation*}%
generate a projective representation of the symplectic group. We then show
that if the parameter $\nu $ is chosen to be index defined in \S 3, then
these operators generate $\limfunc{Mp}(Z,\sigma )$.

\item In \S 5 we construct a phase-space Weyl calculus along the lines
indicated above; that calculus is symplectically covariant with respect to
conjugation with the metaplectic operators of \S 4: an immediate
generalization of a deep result of Shale shows that this covariance actually
characterizes uniquely the Weyl operators we have constructed.
\end{itemize}

Let us precise some notations that will be used throughout this paper; we
take the opportunity to recall some basic results.

\subsubsection*{Symplectic notations}

Let $(E,\omega)$ be a finite-dimensional symplectic space; we denote by $%
\limfunc{Sp}(E,\omega)$ and $\func{Lag}(E,\omega)$ the symplectic group and
Lagrangian Grassmannian and by 
\begin{align*}
\pi^{\limfunc{Sp}} & :\limfunc{Sp}\nolimits_{\infty}(E,\omega)\longrightarrow%
\limfunc{Sp}(E,\omega) \\
\pi^{\func{Lag}} & :\func{Lag}_{\infty}(E,\omega )\longrightarrow\func{Lag}%
(E,\omega)
\end{align*}
the corresponding universal coverings. We will call $\func{Lag}%
_{\infty}(E,\omega)$ the \textquotedblleft Maslov bundle\textquotedblright\
of the symplectic space $(E,\omega)$.

Let $X=\mathbb{R}^{n}$; the standard symplectic structure on $Z=X\oplus
X^{\ast}$ is defined by%
\begin{equation*}
\sigma(z,z^{\prime})=\left\langle p,x^{\prime}\right\rangle -\left\langle
p^{\prime},x\right\rangle \text{ \ for \ }z=(x,p)\text{ , }z^{\prime
}=(x^{\prime},p^{\prime})\text{.}
\end{equation*}
Identifying $Z$ with $\mathbb{R}^{2n}$ we have $\sigma(z,z^{\prime
})=\left\langle Jz,z^{\prime}\right\rangle $ where $\left\langle z,z^{\prime
}\right\rangle =\left\langle x,x^{\prime}\right\rangle +\left\langle
p,p^{\prime\prime}\right\rangle $ is the usual Euclidean scalar product on $%
\mathbb{R}^{2n}$ and $J=%
\begin{bmatrix}
0 & I \\ 
-I & 0%
\end{bmatrix}
$. The subgroup $\limfunc{Sp}(Z,\sigma)\cap\limfunc{O}(2n,\mathbb{R})$ is
identified with the unitary group $\limfunc{U}(n,\mathbb{C})$ by the mapping%
\begin{equation*}
\iota:%
\begin{bmatrix}
A & -B \\ 
B & B%
\end{bmatrix}
\longmapsto A+iB\text{;}
\end{equation*}
the action of $\limfunc{U}(n,\mathbb{C})$ on $\func{Lag}(Z,\sigma)$ is
denoted by $(u,\ell)\longmapsto u\ell$.

\subsubsection*{Maslov index on $\limfunc{Sp}(Z,\protect\sigma)$}

The Maslov index for loops in $\limfunc{Sp}(Z,\sigma)$ is defined as
follows: let $\gamma:[0,1]\longrightarrow\limfunc{Sp}(Z,\sigma)$ be such
that $\gamma(0)=\gamma(1)$, and set $\gamma(t)=S_{t}$. Then $%
U_{t}=(S_{t}S_{t})^{-1/2}S_{t}$ is the orthogonal part in the polar
decomposition of $S_{t}$: 
\begin{equation*}
U_{t}\in\limfunc{Sp}(Z,\sigma)\cap\limfunc{O}(2n,\mathbb{R})\text{;}
\end{equation*}
let us denote by $u_{t}$ its image $\iota(U_{t})$ in $\limfunc{U}(n,\mathbb{C%
})$ and define $\rho(S_{t})=\det u_{t}$. The Maslov index of $\gamma$ is the
degree of the loop $t\longmapsto\rho(S_{t})$ in $S^{1}$:%
\begin{equation*}
m(\gamma)=\deg[t\longmapsto\det(\iota(U_{t}))\text{ , }0\leq t\leq1]\text{.}
\end{equation*}

\subsubsection*{Generalized Fresnel integral}

We will need the following Fresnel-type formula: Let $F$ be the Fourier
transform on $\mathbb{R}^{m}$%
\begin{equation*}
Ff(v)=\left( \tfrac{1}{2\pi}\right) ^{m/2}\int_{\mathbb{R}%
^{m}}e^{-i\left\langle v,u\right\rangle }f(u)du\text{;}
\end{equation*}
if $M$ is a real symmetric $m\times m$ matrix such that $M>0$ and $f:$ $%
u\longmapsto e^{\frac{i}{2}\left\langle Mu,u\right\rangle }$ then we
have-the Fresnel-type formula%
\begin{equation}
Ff(v)=|\det M|^{-1/2}e^{\frac{i\pi}{4}\limfunc{sign}M}e^{-\frac{i}{2}%
\left\langle M^{-1}v,v\right\rangle }  \label{fres}
\end{equation}
where $\limfunc{sign}M$, the \textquotedblleft signature\textquotedblright\
of $M$, is the number of $>0$ eigenvalues of $M$ minus the number of $<0$
eigenvalues.

\subsubsection*{Weyl--Wigner--Moyal formalism}

We refer to the standard literature (for instance \cite%
{Folland,Gro,Stein,Wong}) for detailed studies of Weyl pseudo-differential
calculus and of the related Weyl--Wigner--Moyal formalism. The Wigner--Moyal
transform $W(f,g)$ of $f,g\in\mathcal{S}(X)$ is defined by 
\begin{equation}
W(f,g)(x,p)=\left( \tfrac{1}{2\pi}\right) ^{n}\int_{X}e^{-i\left\langle
p,y\right\rangle }f(x+\tfrac{1}{2}y)\overline{g}(x-\tfrac{1}{2}y)dy\text{;}
\label{wm}
\end{equation}
it extends to a mapping%
\begin{equation*}
W:\mathcal{S}(X)\times\mathcal{S}^{\prime}(X)\longrightarrow\mathcal{S}%
^{\prime}(X)\text{.}
\end{equation*}
The Weyl operator $\widehat{A}$ with \textquotedblleft
symbol\textquotedblright\ $a\in\mathcal{S}^{\prime}(X)$ is defined by%
\begin{equation*}
\left\langle \widehat{A}f,\phi\right\rangle =\left\langle a,W(f,\overline {%
\phi})\right\rangle
\end{equation*}
for $f,g\in\mathcal{S}(X)$; $\left\langle \cdot,\cdot\right\rangle $ denotes
the usual distributional bracket. The symplectic Fourier transform of $a\in%
\mathcal{S}(Z)$ is defined by%
\begin{equation*}
\mathcal{F}_{\sigma}a(z)=f_{\sigma}(z)=\int_{Z}e^{-i\sigma(z,z^{%
\prime})}a(z^{\prime})dz^{\prime}
\end{equation*}
and extends to $\mathcal{S}^{\prime}(Z)$; setting $a_{\sigma}=\mathcal{F}%
_{\sigma}a$ (the twisted symbol) we have%
\begin{equation*}
\widehat{A}f(x)=\left( \tfrac{1}{2\pi}\right) ^{n}\int_{Z}a_{\sigma }(z)%
\widehat{T}(z)f(x)dz
\end{equation*}
(interpreted in the distributional sense). Let $a$ and $b$ be the symbols of 
$A$ and $B$ respectively; then the twisted symbol $c_{\sigma}$ of the
compose $C=AB$ (when defined) is given by the \textquotedblleft twisted
convolution\textquotedblright\ 
\begin{equation*}
c_{\sigma}(z)=\left( \tfrac{1}{2\pi}\right) ^{n}\int_{Z}e^{\frac{i}{2}%
\sigma(z,z^{\prime})}a_{\sigma}(z-z^{\prime})b_{\sigma}(z^{\prime})dz^{%
\prime }\text{.}
\end{equation*}

\section{The ALM and Maslov Indices}

We review, without proofs, the main formulas and results developed in \cite%
{JMPA,Wiley}; for an alternative construction due to Dazord see \cite{Dazord}%
. In \cite{CLM} Cappell \textit{et al.} compare the ALM index to various
other indices used in mathematics. We begin by defining a notion of
signature for triples of Lagrangian planes (it is sometimes called
\textquotedblleft Maslov triple index\textquotedblright).

\subsection{The Kashiwara--Wall signature}

For proofs see \cite{CLM,LV,Wiley}. Let $(E,\omega )$ be a symplectic space, 
$\dim E=n<\infty $. Let $(\ell ,\ell ^{\prime },\ell ^{\prime \prime })$ be
a triple of elements of $\func{Lag}(E,\omega )$. By definition the
Kashiwara--Wall signature (or index) \cite{LV,Wall} of that triple is the
signature, denoted $\tau (\ell ,\ell ^{\prime },\ell ^{\prime \prime })$, of
the quadratic form%
\begin{equation*}
(z,z^{\prime },z^{\prime \prime })\longmapsto \omega (z,z^{\prime })+\omega
(z^{\prime },z^{\prime \prime })+\omega (z^{\prime \prime },z)
\end{equation*}%
on $\ell \oplus \ell ^{\prime }\oplus \ell ^{\prime \prime }$. The kernel of
that quadratic form is isomorphic to $(\ell \cap \ell ^{\prime })\oplus
(\ell ^{\prime }\cap \ell ^{\prime \prime })\oplus (\ell ^{\prime \prime
}\cap \ell )$ hence%
\begin{equation*}
\tau (\ell ,\ell ^{\prime },\ell ^{\prime \prime })\equiv n+\dim \ell \cap
\ell ^{\prime }+\dim \ell ^{\prime }\cap \ell ^{\prime \prime }+\dim \ell
^{\prime \prime }\cap \ell \text{ \ }\func{mod}2\text{.}
\end{equation*}

The Kashiwara--Wall signature has the following properties:\bigskip

\noindent\fbox{K.1} $\tau$\textit{\ is antisymmetric}:%
\begin{equation*}
\tau(\mathfrak{p}(\ell,\ell^{\prime},\ell^{\prime\prime}))=(-1)^{\mathrm{sgn}%
(\mathfrak{p})}\tau(\ell,\ell^{\prime},\ell^{\prime\prime})
\end{equation*}
for any permutation $\mathfrak{p}$ of the set $\{\ell,\ell^{\prime},\ell^{%
\prime\prime}\}$; $\mathrm{sgn}(\mathfrak{p})=0$ if $\mathfrak{p}$ is even, $%
1$ if $\mathfrak{p}$ is odd. In particular $\tau(\ell,\ell^{\prime
},\ell^{\prime\prime})=0$ if any two of the three Lagrangian planes $\ell
,\ell^{\prime},\ell^{\prime\prime}$ are identical;\bigskip

\noindent\fbox{K.2} $\tau$\textit{\ is }$\limfunc{Sp}(E,\omega )$\textit{%
-invariant}:%
\begin{equation*}
\tau(S\ell,S\ell^{\prime},S\ell^{\prime\prime})=\tau(\ell,\ell^{\prime},%
\ell^{\prime\prime})
\end{equation*}
for every $S\in\limfunc{Sp}(E,\omega)$;\bigskip

\noindent\fbox{K.3} $\tau$\textit{\ }is\textit{\ locally constant }on each
set set of triples 
\begin{equation*}
\{(\ell,\ell^{\prime},\ell^{\prime\prime}):\dim\ell\cap\ell^{\prime}=k;\text{
}\dim\ell^{\prime}\cap\ell^{\prime\prime}=k^{\prime};\text{ }\dim\ell
^{\prime\prime}\cap\ell=k^{\prime\prime}\}
\end{equation*}
where $0\leq k,k^{\prime},k^{\prime\prime}\leq n$;\bigskip

\noindent\fbox{K.4} $\tau$\textit{\ is a cocycle}:%
\begin{equation}
\tau(\ell,\ell^{\prime},\ell^{\prime\prime})-\tau(\ell^{\prime},\ell
^{\prime\prime},\ell^{\prime\prime\prime})+\tau(\ell^{\prime},\ell
^{\prime\prime},\ell^{\prime\prime\prime})-\tau(\ell^{\prime},\ell
^{\prime\prime},\ell^{\prime\prime\prime})=0  \label{coka}
\end{equation}
for all $\ell,\ell^{\prime},\ell^{\prime\prime},\ell^{\prime\prime\prime}$
in $\func{Lag}(E,\omega)$.\bigskip

\noindent \fbox{K.5} $\tau $\textit{\ is dimensionally additive}: Let $%
(E,\omega )=(E^{\prime }\oplus E^{\prime \prime },\omega ^{\prime }\oplus
\omega ^{\prime \prime })$. Identifying $\func{Lag}(E^{\prime },\omega
^{\prime })\oplus \func{Lag}(E^{\prime \prime },\omega ^{\prime \prime })$
with a subset of $\func{Lag}(E,\omega )$ we have 
\begin{equation}
\tau (\ell _{1}^{\prime }\oplus \ell _{1}^{\prime \prime },\ell _{2}^{\prime
}\oplus \ell _{2}^{\prime \prime },\ell _{3}^{\prime }\oplus \ell
_{3}^{\prime \prime })=\tau ^{\prime }(\ell _{1}^{\prime },\ell _{2}^{\prime
},\ell _{3}^{\prime })+\tau ^{\prime \prime }(\ell _{1}^{\prime \prime
},\ell _{2}^{\prime \prime },\ell _{3}^{\prime \prime })  \label{tausomme}
\end{equation}%
where $\tau ^{\prime }$ and $\tau ^{\prime \prime }$ are the Kashiwara--Wall
signatures on $\func{Lag}(E^{\prime },\omega ^{\prime })$ and $\func{Lag}%
(E^{\prime \prime },\omega ^{\prime \prime })$ and $\tau =\tau ^{\prime
}\oplus \tau ^{\prime \prime }$ that on $\func{Lag}(E,\omega )$.\bigskip

In addition to these fundamental properties which characterize $\tau $, the
Kashiwara--Wall signature enjoys the following subsidiary properties which
are very useful for practical calculations:\bigskip

\noindent\fbox{K.6} If $\ell\cap\ell^{\prime\prime}=0$ then $\tau(\ell
,\ell^{\prime},\ell^{\prime\prime})$ is the signature of the quadratic form 
\begin{equation*}
Q^{\prime}(z^{\prime})=\omega(P_{\ell\ell^{\prime\prime}}z^{\prime},z^{%
\prime })=\omega(z^{\prime},P_{\ell^{\prime\prime}\ell}z^{\prime})
\end{equation*}
on $\ell^{\prime},$ where $P_{\ell\ell^{\prime\prime}}$ is the projection
onto $\ell$ along $\ell^{\prime\prime}$ and $P_{\ell^{\prime\prime}%
\ell}=I-P_{\ell\ell^{\prime\prime}}$ is the projection on $%
\ell^{\prime\prime}$ along $\ell$.\bigskip

\noindent\fbox{K.7} Let $(\ell,\ell^{\prime},\ell^{\prime\prime})$ be a
triple of Lagrangian planes such that an $\ell=\ell\cap\ell^{\prime}+\ell%
\cap \ell^{\prime\prime}$. Then $\tau(\ell,\ell^{\prime},\ell^{\prime%
\prime})=0$.\bigskip

\noindent\fbox{K.8} Let $(E,\omega)$ be the standard symplectic space $%
(X\oplus X^{\ast},\sigma)$. Let $\ell_{A}=\{(x,Ax)$, $x\in X\}$ where $A$ is
a symmetric linear mapping $X\longrightarrow X^{\ast}$. Then 
\begin{equation}
\tau(X^{\ast},\ell_{A},X)=\limfunc{sign}(A)\text{.}  \label{4triple1}
\end{equation}

\begin{remark}
It is proven in \cite{CLM} that the properties K.1, K.2, K.5 uniquely
characterize the Kashiwara--Wall signature $\tau $ up to a factor. Property
K.8 then appears as a \textquotedblleft normalization
property\textquotedblright\ determining unambiguously $\tau $.
\end{remark}

The Kashiwara--Wall signature is related to various other algebraic objects
appearing in the literature. Here are two examples; for more see \cite{CLM}
where, for instance, the relationship between $\tau $ and Wall's \cite{Wall}
original index is investigated.

\begin{example}
\label{exun}In \cite{Leray} Leray defined the index of inertia $\func{Inert}%
(\ell,\ell^{\prime},\ell^{\prime\prime})$ of a triple of pairwise transverse
elements of $\func{Lag}(E,\omega)$ as being the common index of inertia of
the three quadratic forms $z\longmapsto \omega(z,z^{\prime})$, $%
z^{\prime}\longmapsto\omega(z^{\prime},z^{\prime \prime})$, $%
z^{\prime\prime}\longmapsto\omega(z^{\prime\prime},z^{\prime})$ where $%
(z,z^{\prime},z^{\prime\prime})\in\ell\times\ell^{\prime}\times
\ell^{\prime\prime}$ is such that $z+z^{\prime}+z^{\prime\prime}=0$. It
easily follows from property (K.6) of $\tau$ that 
\begin{equation*}
\tau(\ell,\ell^{\prime},\ell^{\prime\prime})=2\func{Inert}(\ell
,\ell^{\prime},\ell^{\prime\prime})-n\text{.}
\end{equation*}
\end{example}

\begin{example}
\label{exdeux}In \cite{RS} Robbin and Salamon's define a \textquotedblleft
composition form\textquotedblright\ $Q$ for pairs $(S,S^{\prime})$ of
elements of $\func{Sp}(Z,\sigma)$ such that $SX^{\ast}\cap
X^{\ast}=S^{\prime }X^{\ast}\cap X^{\ast}=0$; it is given by 
\begin{equation*}
Q(S,S^{\prime})=\limfunc{sign}(B^{-1}B^{\prime\prime}(B^{\prime})^{-1})
\end{equation*}
when 
\begin{equation*}
S=%
\begin{bmatrix}
A & B \\ 
C & D%
\end{bmatrix}
\text{ \ , \ }S^{\prime}=%
\begin{bmatrix}
A^{\prime} & B^{\prime} \\ 
C^{\prime} & D^{\prime}%
\end{bmatrix}
\text{ \ , \ }S^{\prime\prime}=%
\begin{bmatrix}
A^{\prime\prime} & B^{\prime\prime} \\ 
C^{\prime\prime} & D^{\prime\prime}%
\end{bmatrix}
\text{.}
\end{equation*}
We have shown in \cite{MSDG2} that: 
\begin{equation}
Q(S,S^{\prime})=\tau(X^{\ast},SX^{\ast},SS^{\prime}X^{\ast})\text{.}
\label{qtau}
\end{equation}
\end{example}

\subsection{The ALM index}

We denote by $\alpha$ and $\beta$ the generators with index $0$ of $\pi _{1}[%
\func{Lag}(E,\omega)]\simeq(\mathbb{Z},+)$ and $\pi _{1}[\limfunc{Sp}%
(E,\omega)]\simeq(\mathbb{Z},+)$, respectively. Assume that $%
(E,\omega)=(Z,\sigma)$ and identify $(x,p)$ with the vector $%
(x_{1},p_{1}...,x_{n},p_{n})$. The direct sum 
\begin{equation*}
\func{Lag}(1)\otimes\cdots\oplus\func{Lag}(1)\text{ (}n\text{ terms)}
\end{equation*}
is identified with a subset of $\func{Lag}(Z,\sigma)$. Consider the loop $%
\beta_{(1)}:t\longmapsto e^{2\pi it}$, $0\leq t\leq1$, in $W(1,\mathbb{C}%
)\equiv\func{Lag}(1)$. Then $\beta=\beta_{(1)}\oplus I_{2n-2}$ where $%
I_{2n-2}$ is the identity in $W(n-1,\mathbb{C})$. Similarly, denoting by $%
\limfunc{Sp}(1)$ the symplectic group acting on pairs $(x_{j},p_{j})$ the
direct sum%
\begin{equation*}
\limfunc{Sp}(1)\oplus\limfunc{Sp}(1)\oplus\cdots\otimes \limfunc{Sp}(1)\text{
\ (}n\text{ terms})
\end{equation*}
is identified with a subgroup of $\limfunc{Sp}(Z,\sigma)$. Let $J_{1}=%
\begin{bmatrix}
0 & 1 \\ 
-1 & 0%
\end{bmatrix}
$. Then $\alpha$ is identified with%
\begin{equation}
\alpha:t\longmapsto e^{2\pi tJ_{1}}\oplus I_{n-2}\text{ \ , \ }0\leq t\leq1
\label{sploop}
\end{equation}
where $I_{2n-2}$ is the identity on $\mathbb{R}^{2n-2}$.

The Arnol'd--Leray--Maslov (for short: \textquotedblleft\emph{ALM}%
\textquotedblright)\emph{\ }index on $(E,\omega)$ is the unique mapping%
\begin{equation*}
\func{Lag}_{\infty}(E,\omega)\times\func{Lag}_{\infty
}(E,\omega)\longrightarrow\mathbb{Z}
\end{equation*}
having the following characteristic property:\bigskip

\noindent\fbox{ALM.1} \textit{Topological and cocycle condition}: $\mu$ is
locally constant on the sets%
\begin{equation}
\{(\ell_{\infty},\ell_{\infty}^{\prime}):\dim\ell\cap\ell^{\prime}=k\}
\label{sets}
\end{equation}
($0\leq k\leq n$) and satisfies 
\begin{equation}
\mu(\ell_{\infty},\ell_{\infty}^{\prime})-\mu(\ell_{\infty},\ell_{\infty
}^{\prime\prime})+\mu(\ell_{\infty}^{\prime},\ell_{\infty}^{\prime\prime
})=\tau(\ell,\ell^{\prime},\ell^{\prime\prime})\text{.}  \label{un}
\end{equation}

The ALM\ index has the following additional properties:\bigskip

\noindent\fbox{ALM.2} \textit{Antisymmetry}:%
\begin{equation}
\mu(\ell_{\infty},\ell_{\infty}^{\prime})=-\mu(\ell_{\infty}^{\prime},\ell_{%
\infty})\text{ \ , \ \ }\mu(\ell_{\infty},\ell_{\infty})=0  \label{anti}
\end{equation}

\noindent\fbox{ALM.3} \textit{Value modulo} $2$: We have%
\begin{equation}
\mu(\ell_{\infty},\ell_{\infty}^{\prime})\equiv n+\dim\ell\cap\ell^{\prime }%
\text{ \ \ }\func{mod}2\text{.}  \label{mod2}
\end{equation}

\noindent\fbox{ALM.4} \textit{Action of }$\pi_{1}[\func{Lag}(E,\omega)]$: we
have%
\begin{equation}
\mu(\beta^{r}\ell_{\infty},\beta^{r^{\prime}}\ell_{\infty}^{\prime})=\mu
(\ell_{\infty},\ell_{\infty}^{\prime})+2(r-r^{\prime})  \label{trois}
\end{equation}
for all integers $r$ and $r^{\prime}$.\bigskip

\noindent(In particular $\mu(\beta^{r}\ell_{\infty},\ell_{\infty})$ is twice
the Maslov index of any Lagrangian loop homeomorphic to $\beta^{r}$.)\bigskip

\noindent\fbox{ALM.5} \textit{Dimensional additivity}: Let $%
E=E^{\prime}\oplus E^{\prime\prime}$ and $\omega=\omega^{\prime}\oplus%
\omega^{\prime}$. If $\mu^{\prime}$ and $\mu^{\prime\prime}$ are the ALM
indices on $\func{Lag}_{\infty}(E^{\prime},\omega^{\prime})$, $\func{Lag}%
_{\infty}(E^{\prime\prime},\omega^{\prime\prime})$ then 
\begin{equation}
\mu(\ell_{1,\infty}^{\prime}\oplus\ell_{1,\infty}^{\prime\prime},\ell_{2,%
\infty}^{\prime}\oplus\ell_{2,\infty}^{\prime\prime})=\mu^{\prime
}(\ell_{1,\infty}^{\prime},\ell_{2,\infty}^{\prime})+\mu^{\prime\prime}(%
\ell_{1,\infty}^{\prime\prime},\ell_{2,\infty}^{\prime\prime})\text{.}
\label{muso}
\end{equation}
\medskip

The natural action 
\begin{equation*}
\limfunc{Sp}(E,\omega)\times\func{Lag}(E,\omega)\longrightarrow \func{Lag}%
(E,\omega)
\end{equation*}
induces an action 
\begin{equation*}
\limfunc{Sp}\nolimits_{\infty}(E,\omega)\times\func{Lag}_{\infty}(E,\omega)%
\longrightarrow\func{Lag}_{\infty}(E,\omega)
\end{equation*}
such that%
\begin{equation}
S_{\infty}(\beta^{2}\ell_{\infty})=(\alpha S_{\infty})\ell_{\infty}=\beta
^{2}(S_{\infty}\ell_{\infty})  \label{deux}
\end{equation}
where $\alpha$ (resp. $\beta$) are the generators of $\pi_{1}[\limfunc{Sp}%
(E,\omega)]$ and $\pi_{1}[\func{Lag}(E,\omega)]$ previously defined. The
uniqueness of an index satisfying property (ALM.1) together with the
symplectic invariance (K.2) of $\tau$ imply that:\bigskip

\noindent\fbox{ALM.6} \textit{Symplectic invariance}: for all $S_{\infty}\in%
\limfunc{Sp}_{\infty}(E,\omega)$ we have 
\begin{equation}
\mu(S_{\infty}\ell_{\infty},S_{\infty}\ell_{\infty}^{\prime})=\mu(\ell
_{\infty},\ell_{\infty}^{\prime})\text{.}  \label{quatre}
\end{equation}
\bigskip

Let us give a procedure for calculating explicitly the ALM\ index.

Assume that $(E,\omega)$ is the standard symplectic space $(X\oplus X^{\ast
},\sigma)$. Identifying $\func{Lag}(Z,\sigma)$ with the set%
\begin{equation*}
W(n,\mathbb{C})=\{w\in\limfunc{U}(n,\mathbb{C}):w=w^{T}\}
\end{equation*}
using the mapping which to $\ell=uX^{\ast}$ ($u\in\limfunc{U}(n,\mathbb{C})$%
) associates $w=uu^{T},$ the Maslov bundle $\func{Lag}_{\infty}(Z,\sigma) $
is identified with%
\begin{equation*}
W_{\infty}(n,\mathbb{C})=\{(w,\theta):w\in W(n,\mathbb{C})\text{, }\det
w=e^{i\theta}\};
\end{equation*}
the projection $\pi^{\func{Lag}}:\ell_{\infty}\longmapsto\ell$ becomes $%
(w,\theta)\longmapsto w$. The ALM\ index is then calculated as follows:

\begin{itemize}
\item If $\ell\cap\ell^{\prime}=0$ then 
\begin{equation}
\mu(\ell_{\infty},\ell_{\infty}^{\prime})=\frac{1}{\pi}\left[ \theta
-\theta^{\prime}+i\limfunc{Tr}\func{Log}(-w(w^{\prime})^{-1}\right]
\label{souriau}
\end{equation}
(the transversality condition $\ell\cap\ell^{\prime}$ is equivalent to $%
-w(w^{\prime})^{-1}$ having no negative eigenvalue);

\item If $\ell\cap\ell^{\prime}\neq0$ one chooses $\ell^{\prime\prime}$ such
that $\ell\cap\ell^{\prime\prime}=\ell^{\prime}\cap\ell^{\prime\prime}=0$
and one then calculates $\mu(\ell_{\infty},\ell_{\infty}^{\prime})$ using
the formula (\ref{un}) the values of $\mu(\ell_{\infty},\ell_{\infty}^{%
\prime\prime})$ and $\mu(\ell_{\infty}^{\prime},\ell_{\infty}^{\prime\prime
})$ given by (\ref{souriau}). ($\mu(\ell_{\infty},\ell_{\infty}^{\prime})$
does not depend on the choice of $\ell^{\prime\prime}$ in view of the
cocycle property (\ref{coka}) of $\tau$.)\medskip
\end{itemize}

The ALM index is useful for expressing in a simple way various Lagrangian
path intersection indices. For instance, in \cite{RS} is defined an
intersection index for paths in $\func{Lag}(Z,\sigma)$ with arbitrary
endpoints, counting algebraically the intersections of a path $\Lambda$ in $%
\func{Lag}(Z,\sigma)$ with the caustic $\Sigma_{\ell}=\{\ell^{\prime
}:\ell\cap\ell^{\prime}=0\}$:

\begin{example}
\label{ex1}Let $\mu_{\text{RS}}$ be the Robbin--Salamon index defined in 
\cite{RS}. That index associates to a continuous path $\Lambda
:[a,b]\longrightarrow\func{Lag}(Z,\sigma)$ and $\ell\in \func{Lag}(Z,\sigma)$
a number $\mu_{\text{RS}}(\Lambda,\ell)$. In \cite{MSDG1,MSDG2} we have
shown that 
\begin{equation}
\mu_{\text{RS}}(\Lambda,\ell)=\frac{1}{2}(m(\ell_{b,\infty},\ell_{\infty
})-m(\ell_{a,\infty},\ell_{\infty}))  \label{foufoun}
\end{equation}
where $\ell_{\infty}$ is an arbitrary element of $\func{Lag}_{\infty
}(Z,\sigma)$ covering $\ell$; $\ell_{a,\infty}$ is the equivalence class of
an arbitrary path $\Lambda_{0a}$ joining the chosen base point $\ell_{0} $
of $\func{Lag}(Z,\sigma)$ to $\ell_{a}=\Lambda(a)$, and $\ell_{b,\infty}$ is
the equivalence class of the concatenation $\Lambda_{0a}\ast\Lambda$.
\end{example}

The theory of that index has been applied and extended with success to
problems in functional analysis \cite{BF1} and in Morse theory where it
provides useful \textquotedblleft spectral flow\textquotedblright\ formulas
(see Piccione and his collaborators \cite{GPP,Piccione2}).

\subsection{Relative Maslov indices on $\limfunc{Sp}(Z,\protect\sigma)$}

The \emph{Maslov indices} $\mu_{\ell}$ on $\limfunc{Sp}_{\infty }(Z,\sigma)$
are defined in terms of the ALM index as follows. Let $\ell_{\infty}\in\func{%
Lag}_{\infty}(Z,\sigma)$ and $S_{\infty}\in\limfunc{Sp}_{\infty}(Z,\sigma)$;
formulae (\ref{deux}), (\ref{trois}) imply that the integer $%
\mu(S_{\infty}\ell_{\infty},\ell _{\infty})$ only depends on $\ell=\pi^{%
\func{Lag}}(\ell_{\infty})$. The \textquotedblleft Maslov index on $\limfunc{%
Sp}_{\infty}(Z,\sigma)$ relative to $\ell$\textquotedblright\ is the mapping 
$\mu_{\ell}:\limfunc{Sp}_{\infty}(Z,\sigma)\longrightarrow\mathbb{Z}$
defined by%
\begin{equation}
\mu_{\ell}(S_{\infty})=\mu(S_{\infty}\ell_{\infty},\ell_{\infty})\text{.}
\label{cinq}
\end{equation}
It follows from the cocycle property (\ref{un}) in (ALM.1) that:\bigskip

\noindent\fbox{M.1} \textit{Uniqueness and product}: $\mu_{\ell}$ is the
only mapping $\limfunc{Sp}_{\infty}(Z,\sigma)\longrightarrow\mathbb{Z}$
which is locally constant on each set 
\begin{equation}
\limfunc{Sp}\nolimits_{\ell}(n;k)=\{S\in\limfunc{Sp}(Z,\sigma):\dim(S\ell%
\cap\ell)=k\}  \label{spnk}
\end{equation}
($0\leq k\leq n$) and such that 
\begin{equation}
\mu_{\ell}(S_{\infty}S_{\infty}^{\prime})=\mu_{\ell}(S_{\infty})+\mu_{\ell
}(S_{\infty}^{\prime})+\tau(\ell,S\ell,SS^{\prime}\ell)\text{.}  \label{uno}
\end{equation}
\medskip

\noindent \fbox{M.2} \textit{Antisymmetry}: for all $S_{\infty }\in \limfunc{%
Sp}_{\infty }(Z,\sigma )$%
\begin{equation}
\mu _{\ell }(S_{\infty }^{-1})=-\mu _{\ell }(S_{\infty })\text{ \ \ , \ \ }%
\mu _{\ell }(I_{\infty })=0  \label{zero}
\end{equation}%
($I_{\infty }$ the unit of $\limfunc{Sp}_{\infty }(Z,\sigma )$);\bigskip

\noindent\fbox{M.3}\textit{\ Action of} $\pi_{1}[\limfunc{Sp}(Z,\sigma)]$:
let $\alpha$ be the generator of $\pi_{1}[\limfunc{Sp}(Z,\sigma)]$; the 
\begin{equation}
\mu_{\ell}(\alpha^{r}S_{\infty})=\mu_{\ell}(S_{\infty})+4r  \label{sex}
\end{equation}
for all $S_{\infty}\in\limfunc{Sp}_{\infty}(Z,\sigma)$ and $r\in\mathbb{Z}$%
.\bigskip

\noindent \fbox{M.4} \textit{Dimensional additivity}: Let $Z^{\prime }=%
\mathbb{R}^{2n^{\prime },}$, $Z^{\prime \prime }=\mathbb{R}^{2n^{\prime
\prime }}$, $n^{\prime }+n^{\prime }=n$. Identifying $\limfunc{Sp}(Z^{\prime
},\sigma ^{\prime })\oplus \limfunc{Sp}(Z^{\prime \prime },\sigma ^{\prime
\prime })$ with a subgroup of $\limfunc{Sp}(Z,\sigma )$ we have 
\begin{equation*}
\mu _{\ell ^{\prime }\oplus \ell ^{\prime \prime }}(S_{\infty }^{\prime
}\oplus S_{\infty }^{\prime \prime })=\mu _{\ell ^{\prime }}(S_{\infty
}^{\prime })+\mu _{\ell ^{\prime \prime }}(S_{\infty }^{\prime \prime })%
\text{.}
\end{equation*}

Notice that it follows from formula (\ref{mod2}) that%
\begin{equation}
\mu_{\ell}(S_{\infty})\equiv n+\dim(S\ell\cap\ell)\text{ \ \ }\func{mod}2
\label{mumod2}
\end{equation}

Following formula, which immediately follows from the cocycle property (K.4)
of $\tau$, describes the behavior of the Maslov index under changes of $\ell$%
:

\begin{align}
\mu_{\ell}(S_{\infty})-\mu_{\ell^{\prime}}(S_{\infty}) & =\tau(S\ell
,\ell,\ell^{\prime})-\tau(S\ell,S\ell^{\prime},\ell^{\prime})  \label{mule}
\\
& =\tau(S\ell,\ell,S\ell^{\prime})-\tau(\ell,S\ell^{\prime},\ell^{\prime })%
\text{.}  \notag
\end{align}

It is sometimes advantageous to work with the \textquotedblleft reduced
Maslov index\textquotedblright\ relative to $\ell \in \func{Lag}(Z,\sigma )$%
; it is the function $m_{\ell }:\limfunc{Sp}_{\infty }(Z,\sigma
)\longrightarrow \mathbb{Z}$ defined by%
\begin{equation*}
m_{\ell }(S_{\infty })=m(S_{\infty }\ell _{\infty },\ell _{\infty })=\frac{1%
}{2}(\mu _{\ell }(S_{\infty })+n+\dim (S\ell \cap \ell ))\text{.}
\end{equation*}%
Notice that in view of (\ref{mumod2}) we have%
\begin{equation*}
m_{\ell }(S_{\infty })\equiv n+\dim (S\ell \cap \ell )\text{ \ }\func{mod}2%
\text{.}
\end{equation*}

The properties of the reduced index $m_{\ell}$ are immediately deduced from
those of $\mu_{\ell}$; for instance%
\begin{equation}
m_{\ell}(S_{\infty}S_{\infty}^{\prime})=m_{\ell}(S_{\infty})+m_{\ell
}(S_{\infty}^{\prime})+\limfunc{Inert}(\ell,S\ell,SS^{\prime}\ell)
\label{mredux}
\end{equation}
and%
\begin{equation}
m_{\ell}(\alpha^{r}S_{\infty})=m_{\ell}(S_{\infty})+2r  \label{maf}
\end{equation}
for $r\in\mathbb{Z}$.

Exactly as the ALM index allows an easy construction of Lagrangian path
intersection indices (Example \ref{ex1}) the relative Maslov index allows to
construct symplectic path intersection indices:

\begin{example}
Let $\Sigma$ be a continuous path $[a,b]\longrightarrow\limfunc{Sp}%
(Z,\sigma) $; set $S_{t}=\Sigma(t)$. Let $\ell\in\func{Lag}(Z,\sigma )$. The
intersection index of $\Sigma$ with the subvariety $\{S:S\ell\cap
\ell\neq0\} $ of $\limfunc{Sp}(Z,\sigma)$ is by definition%
\begin{equation*}
\mu(\Sigma,\ell)=\frac{1}{2}(m_{\ell}(S_{b,\infty})-m(S_{a,\infty}))
\end{equation*}
where $S_{a,\infty}$ is the homotopy class in $\limfunc{Sp}(Z,\sigma)$ of an
arbitrary path $\Sigma_{0a}$ joining the identity to $S_{a}$ and $%
S_{b,\infty}$ that of the concatenation $\Sigma_{0a}\ast\Sigma$. Choosing $%
\ell=X^{\ast}$ one obtains the symplectic path intersection studied in \cite%
{RS} (see \cite{MSDG2}).
\end{example}

\section{The index $\protect\nu$ on $\limfunc{Sp}_{\infty}(Z,\protect\sigma)$%
}

We are going to study in some detail an index $\nu :\limfunc{Sp}_{\infty
}(Z,\sigma )\longrightarrow \mathbb{Z}$ which will be fundamental in
defining the correct phase of the Weyl symbol of a metaplectic operator.
That index may be seen as an extension of the Conley--Zehnder index \cite%
{CZ,HWZ,SZ} which plays an important role in the theory of periodic
Hamiltonian orbits and their bifurcations \cite{Ciriza}, and in Floer
homology.

We begin by introducing a notion of Cayley transform for symplectic matrices
(a similar transform has been considered by Howe \cite{Howe}).

\subsection{Symplectic Cayley transform}

Let $S\in \limfunc{Sp}(Z,\sigma )$ be such that $\det (S-I)\neq 0$. We will
call the matrix%
\begin{equation}
M_{S}=\frac{1}{2}J(S+I)(S-I)^{-1}  \label{cayley}
\end{equation}
the \textquotedblleft symplectic Cayley transform of $S$\textquotedblright ;
equivalently,%
\begin{equation}
M_{S}=\frac{1}{2}J+J(S-I)^{-1}.  \label{cayleybis}
\end{equation}

The following Lemma summarizes the main properties of the mapping $%
S\longmapsto M_{S}$.

\begin{lemma}
\label{lemcarton}Let $\limfunc{Sp}_{0}(Z,\sigma )$ be the set of all $S\in 
\limfunc{Sp}(Z,\sigma )$ with $\det (S-I)\neq 0$ and $\limfunc{Sym}_{0}(2n,%
\mathbb{R})$ the set of all real $2n\times 2n$ symmetric matrices $M$ such
that $\det (M-\tfrac{1}{2}J)\neq 0$. (i) The symplectic Cayley transform is
a bijection 
\begin{equation*}
\limfunc{Sp}\nolimits_{0}(Z,\sigma )\longrightarrow \limfunc{Sym}%
\nolimits_{0}(2n,\mathbb{R})
\end{equation*}%
whose inverse is given by the formula 
\begin{equation}
S=(M-\tfrac{1}{2}J)^{-1}(M+\tfrac{1}{2}J)  \label{cayleyprod}
\end{equation}%
if $M=M_{S}$. (ii) The symplectic Cayley transform of the product $%
SS^{\prime }$ is (when defined) given by the formula%
\begin{equation}
M_{SS^{\prime }}=M_{S}+(S^{T}-I)^{-1}J(M_{S}+M_{S^{\prime }})^{-1}J(S-I)^{-1}
\label{mss}
\end{equation}%
and we have%
\begin{equation}
(M_{S}+M_{S^{\prime }})^{-1}=-(S^{\prime }-I)(SS^{\prime }-I)^{-1}(S-I)J.
\label{mess}
\end{equation}%
(iii) The symplectic Cayley transform of $S$ and $S^{-1}$ are related by%
\begin{equation}
M_{S^{-1}}=-M_{S}\text{.}  \label{rathobvious}
\end{equation}
\end{lemma}

We omit the proof since the formulae above are obtained by elementary
algebraic manipulations involving the use of the relations $%
SJS^{T}=S^{T}JS=J $ characterizing symplectic matrices; alternatively it is 
\textit{mutatis mutandis} the same as proof of Howe's \cite{Howe} Cayley
transform for symplectic matrices (also see \cite{Folland}, p. 242--243).

\subsection{Definition of $\protect\nu(S_{\infty})$ and first properties}

We define on $Z\oplus Z$ the symplectic form $\sigma^{\ominus}$ by%
\begin{equation*}
\sigma^{\ominus}(z_{1},z_{2};z_{1}^{\prime},z_{2}^{\prime})=%
\sigma(z_{1},z_{1}^{\prime})-\sigma(z_{2},z_{2}^{\prime})
\end{equation*}
and denote by $\limfunc{Sp}^{\ominus}(2n)$ and $\func{Lag}^{\ominus}(2n)$
the corresponding symplectic group and Lagrangian Grassmannian. Let $%
\mu^{\ominus}$ the \textit{ALM} index on $\func{Lag}_{\infty}^{\ominus}(2n)$
and $\mu_{L}^{\ominus}$ the Maslov index on $\limfunc{Sp}_{\infty}^{%
\ominus}(2n)$ relative to $L\in\func{Lag}^{\ominus}(2n)$.

For $S_{\infty}\in\limfunc{Sp}_{\infty}(Z,\sigma)$ we define 
\begin{equation}
\nu(S_{\infty})=\frac{1}{2}\mu^{\ominus}((I\oplus S)_{\infty}\Delta_{\infty
},\Delta_{\infty})  \label{rscz}
\end{equation}
where $(I\oplus S)_{\infty}$ is the homotopy class in $\func{Sp}%
^{\ominus}(2n)$ of the path 
\begin{equation*}
t\longmapsto\{(z,S_{t}z):z\in Z\}\text{ \ , \ }0\leq t\leq1
\end{equation*}
and $\Delta=\{(z,z):z\in Z\}$ the diagonal of $Z\oplus Z$. Setting $%
S_{t}^{\ominus}=I\oplus S_{t}$ we have $S_{t}^{\ominus}\in\limfunc{Sp}%
^{\ominus}(2n)$ hence formulae (\ref{rscz}) is equivalent to 
\begin{equation}
\nu(S_{\infty})=\frac{1}{2}\mu_{\Delta}^{\ominus}(S_{\infty}^{\ominus})
\label{muis}
\end{equation}
where $\mu_{\Delta}^{\ominus}$ is the Maslov index on $\limfunc{Sp}%
_{\infty}^{\ominus}(2n)$ corresponding to $\Delta\in\func{Lag}^{\ominus}(2n)$%
.

Note that replacing $n$ by $2n$ in the congruence (\ref{mumod2}) we have 
\begin{align*}
\mu^{\ominus}((I\oplus S)_{\infty}\Delta_{\infty},\Delta_{\infty}) &
\equiv\dim((I\oplus S)\Delta,\Delta)\text{ \ }\func{mod}2 \\
& \equiv\dim\func{Ker}(S-I)\text{ \ }\func{mod}2
\end{align*}
and hence%
\begin{equation*}
\nu(S_{\infty})\equiv\frac{1}{2}\dim\func{Ker}(S-I)\text{ \ }\func{mod}1
\end{equation*}
so that $\nu(S_{\infty})$ is always an integer since the eigenvalue $1$ of $%
S $ has even multiplicity.

The index $\nu$ has the following rather straightforward properties:\bigskip

\noindent\fbox{$\nu$.1} \textit{Antisymmetry}: For all $S_{\infty}\in%
\limfunc{Sp}_{\infty}(Z,\sigma)$ we have%
\begin{equation*}
\nu(S_{\infty}^{-1})=-\nu(S_{\infty}).
\end{equation*}
This property immediately follows from the equality $(S_{\infty}^{\ominus
})^{-1}=(I\oplus S^{-1})_{\infty}$ and the antisymmetry of $\mu_{\Delta
}^{\ominus}$.\bigskip

\noindent\fbox{$\nu$.2} \textit{Action of }$\pi_{1}[\limfunc{Sp}(Z,\sigma]$:
For all $r\in\mathbb{Z}$ we have 
\begin{equation*}
\nu(\alpha^{r}S_{\infty})=\nu(S_{\infty})+2r
\end{equation*}

To see this it suffices to observe that to the generator $\alpha$ of $\pi
_{1}[\func{Sp}(Z,\sigma)]$ corresponds the generator $I_{\infty}\oplus\alpha$
of $\pi_{1}[\limfunc{Sp}^{\ominus}(2n)]$; in view property (\ref{sex}) of
the Maslov index it follows that 
\begin{align*}
\nu(\alpha^{r}S_{\infty}) & =\frac{1}{2}\mu_{\Delta}^{\ominus}((I_{\infty
}\oplus\alpha)^{r}S_{\infty}^{\ominus}) \\
& =\frac{1}{2}(\mu_{\Delta}^{\ominus}(S_{\infty}^{\ominus})+4r) \\
& =\nu(S_{\infty})+2r\text{.}
\end{align*}

Let us now prove a formula for the index of a product. This formula will be
instrumental in identifying the twisted Weyl symbol of a metaplectic
operator.\bigskip

\noindent \fbox{$\nu $.3} \textit{Product}. If $S_{\infty }$, $S_{\infty
}^{\prime }$, and $S_{\infty }S_{\infty }^{\prime }$ are such that $\det
(S-I)\neq 0$, $\det (S^{\prime }-I)\neq 0$, and $\det (SS^{\prime }-I)\neq 0$
then%
\begin{equation}
\nu (S_{\infty }S_{\infty }^{\prime })=\nu (S_{\infty })+\nu (S_{\infty
}^{\prime })+\frac{1}{2}\limfunc{sign}(M_{S}+M_{S^{\prime }})  \label{muss}
\end{equation}%
where $M_{S}$ is the symplectic Cayley transform of $S$.\bigskip

In view of the product property (\ref{uno}) applied to $\mu _{\Delta
}^{\ominus }$ we have%
\begin{align*}
\nu (S_{\infty }S_{\infty }^{\prime })& =\nu (S_{\infty })+\nu (S_{\infty
}^{\prime })+\tfrac{1}{2}\tau ^{\ominus }(\Delta ,S^{\ominus }\Delta
,S^{\ominus }S^{\prime \ominus }\Delta ) \\
& =\nu (S_{\infty })+\nu (S_{\infty }^{\prime })-\tfrac{1}{2}\tau ^{\ominus
}(S^{\ominus }S^{\prime \ominus }\Delta ,S^{\ominus }\Delta ,\Delta )
\end{align*}%
where $S^{\ominus }=I\oplus S$, $S^{\prime \ominus }=I\oplus S^{\prime }$
and $\tau ^{\ominus }$ is the signature on the symplectic space $(\mathbb{R}%
^{2n}\oplus \mathbb{R}^{2n},\sigma ^{\ominus })$. The condition $\det
(SS^{\prime }-I)\neq 0$ is equivalent to $S^{\ominus }S^{\prime \ominus
}\Delta \cap \Delta =0$ hence we can apply Property K.6 of the
Kashiwara--Wall index with $\ell =S^{\ominus }S^{\prime \ominus }\Delta $, $%
\ell ^{\prime }=S^{\ominus }\Delta $, and $\ell ^{\prime \prime }=\Delta $.
The projection operator onto $S^{\ominus }S^{\prime \ominus }\Delta $ along $%
\Delta $ is easily found to be%
\begin{equation*}
\Pr\nolimits_{S^{\ominus }S^{\prime \ominus }\Delta ,\Delta }=%
\begin{bmatrix}
(I-SS^{\prime })^{-1} & -(I-SS^{\prime })^{-1}\smallskip \\ 
SS^{\prime }(I-SS^{\prime })^{-1} & -SS^{\prime }(I-SS^{\prime })^{-1}%
\end{bmatrix}%
\end{equation*}%
hence $\tau ^{\ominus }(S^{\ominus }S^{\prime \ominus }\Delta ,S^{\ominus
}\Delta ,\Delta )$ is the signature of the quadratic form%
\begin{equation*}
Q(z)=\sigma ^{\ominus }(\Pr\nolimits_{S^{\ominus }S^{\prime \ominus }\Delta
,\Delta }(z,Sz);(z,Sz));
\end{equation*}%
since $\sigma ^{\ominus }=\sigma \ominus \sigma $ we have%
\begin{align*}
Q(z)& =\sigma ((I-SS^{\prime })^{-1}(I-S)z,z))-\sigma (SS^{\prime
}(I-SS^{\prime })^{-1}(I-S)z,Sz)) \\
& =\sigma ((I-SS^{\prime })^{-1}(I-S)z,z))-\sigma (S^{\prime }(I-SS^{\prime
})^{-1}(I-S)z,z)) \\
& =\sigma ((I-S^{\prime })(I-SS^{\prime })^{-1}(I-S)z,z))\text{.}
\end{align*}%
In view of formula (\ref{mess}) in Lemma \ref{lemcarton} we have 
\begin{equation*}
(I-S^{\prime })(SS^{\prime }-I)^{-1}(I-S)=(M_{S}+M_{S^{\prime }})^{-1}J
\end{equation*}%
and hence 
\begin{equation*}
Q(z)=-\left\langle (M_{S}+M_{S^{\prime }})^{-1}Jz,Jz\right\rangle
\end{equation*}%
so that the signature of $Q$ is thus the same as that of the quadratic form%
\begin{equation*}
Q^{\prime }(z)=-\left\langle (M_{S}+M_{S^{\prime }})^{-1}z,z\right\rangle ;
\end{equation*}%
this is $-\limfunc{sign}(M_{S}+M_{S^{\prime }})$ proving formula (\ref{muss}%
).

The index $\nu$ has in addition the following topological property. Let 
\begin{align*}
\limfunc{Sp}\nolimits^{+}(Z,\sigma) & =\{S\in\limfunc{Sp}(Z,\sigma):%
\det(S-I)>0\} \\
\limfunc{Sp}\nolimits^{-}(Z,\sigma) & =\{S\in\limfunc{Sp}(Z,\sigma):%
\det(S-I)<0\} \\
\limfunc{Sp}\nolimits_{0}(Z,\sigma) & =\limfunc{Sp}(Z,\sigma)\backslash(%
\limfunc{Sp}\nolimits^{+}(Z,\sigma)\cup \limfunc{Sp}\nolimits^{-}(Z,\sigma))%
\text{;}
\end{align*}
the sets $\limfunc{Sp}^{\pm}(Z,\sigma)$ are connected and disjoint. We
have:\bigskip

\noindent \fbox{$\nu $.4} Let $S_{\infty }$ be the homotopy class of a path $%
\Sigma $ in $\limfunc{Sp}(Z,\sigma )$ joining the identity to $S\in \limfunc{%
Sp}\nolimits_{0}(Z,\sigma )$, and let $S^{\prime }\in \limfunc{Sp}(Z,\sigma
) $ be in the same connected component $\limfunc{Sp}\nolimits^{\pm
}(Z,\sigma ) $ as $S$. Then $\nu (S_{\infty }^{\prime })=\nu (S_{\infty })$
where $S_{\infty }^{\prime }$ is the homotopy class in $\limfunc{Sp}%
(Z,\sigma )$ of the concatenation of $\Sigma $ and a path joining $S$ to $%
S^{\prime }$ in $\limfunc{Sp}\nolimits_{0}(Z,\sigma )$.\bigskip

Assume in fact that $S$ and $S^{\prime}$ belong to, say, $\func{Sp}%
^{+}(Z,\sigma)$ and let $\Sigma$ be a symplectic path representing $%
S_{\infty }$ and $t\longmapsto\Sigma^{\prime}(t)$ $0\leq t\leq1$, a path
joining $S$ to $S^{\prime}$. Let $S_{\infty}(t)$ be the homotopy class of $%
\Sigma\ast \Sigma^{\prime}(t)$. We have $\det(S(t)-I)>0$ for all $%
t\in\lbrack0,1]$ hence $S_{\infty}^{\ominus}(t)\Delta\cap\Delta\neq0$ as $t$
varies from $0$ to $1$. It follows from the continuity property (M.1) of the
Maslov index that the function $t\longmapsto\mu_{\Delta}^{\ominus}(S_{%
\infty}^{\ominus}(t))$ is constant, hence%
\begin{equation*}
\mu_{\Delta}^{\ominus}(S_{\infty}^{\ominus})=\mu_{\Delta}^{\ominus}(S_{%
\infty
}^{\ominus}(0))=\mu_{\Delta}^{\ominus}(S_{\infty}^{\ominus}(1))=\mu_{\Delta
}^{\ominus}(S_{\infty}^{\prime\ominus})
\end{equation*}
which was to be proven.

\subsection{Relation between $\protect\nu$ and $\protect\mu_{X^{\ast}}$}

The index $\nu$ can be expressed in simple way in terms of the Maslov index $%
\mu_{X^{\ast}}$ on $\limfunc{Sp}_{\infty}(Z,\sigma)$. The following
technical result will be helpful in establishing this important relation.
Recall that $S\in\limfunc{Sp}(Z,\sigma)$ is said to be \textquotedblleft
free\textquotedblright\ if $SX^{\ast}\cap X^{\ast}=0$; this condition is
equivalent to $\det B\neq0$ when $S$ is identified with the matrix 
\begin{equation}
S=%
\begin{bmatrix}
A & B \\ 
C & D%
\end{bmatrix}
\label{czfree}
\end{equation}
in the canonical basis. The set of all free automorphisms is dense in $%
\limfunc{Sp}(Z,\sigma)$. The quadratic form $W$ on $X\oplus X$ defined by 
\begin{equation*}
W(x,x^{\prime})=\frac{1}{2}\left\langle Px,x\right\rangle -\left\langle
Lx,x^{\prime}\right\rangle +\frac{1}{2}\left\langle Qx^{\prime},x^{\prime
}\right\rangle
\end{equation*}
where $P=DB^{-1}$, $L=B^{-1}$, $Q=B^{-1}A$ then generates $S$ in the sense
that $(x,p)=S(x^{\prime},p^{\prime})$ is equivalent to $p=\partial
_{x}W(x,x^{\prime})$, $p^{\prime}=\partial_{x^{\prime}}W(x,x^{\prime})$.

\begin{lemma}
\label{lemma1}Let $S_{W}\in \limfunc{Sp}(Z,\sigma )$ be given by (\ref%
{czfree}).We have 
\begin{equation}
\det (S_{W}-I)=(-1)^{n}\det B\det (B^{-1}A+DB^{-1}-B^{-1}-(B^{T})^{-1})
\label{bofor1}
\end{equation}%
that is:%
\begin{equation*}
\det (S_{W}-I)=(-1)^{n}\det (L^{-1})\det (P+Q-L-L^{T})\text{.}
\end{equation*}%
In particular the symmetric matrix 
\begin{equation*}
P+Q-L-L^{T}=DB^{-1}+B^{-1}A-B^{-1}-(B^{T})^{-1}
\end{equation*}%
is invertible.
\end{lemma}

\begin{proof}
Since $B$ is invertible we can factorize $S-I$ as%
\begin{equation*}
\begin{bmatrix}
A-I & B \\ 
C & D-I%
\end{bmatrix}
=%
\begin{bmatrix}
0 & B \\ 
I & D-I%
\end{bmatrix}%
\begin{bmatrix}
C-(D-I)B^{-1}(A-I) & 0 \\ 
B^{-1}(A-I) & I%
\end{bmatrix}%
\end{equation*}
and hence%
\begin{align*}
\det(S_{W}-I) & =\det(-B)\det(C-(D-I)B^{-1}(A-I)) \\
& =(-1)^{n}\det B\det(C-(D-I)B^{-1}(A-I))\text{.}
\end{align*}
Since $S$ is symplectic we have $C-DB^{-1}A=-(B^{T})^{-1}$ (cf. Step 3 in
the proof of Proposition \ref{propconca}) and hence%
\begin{equation*}
C-(D-I)B^{-1}(A-I))=B^{-1}A+DB^{-1}-B^{-1}-(B^{T})^{-1}\text{;}
\end{equation*}
the Lemma follows
\end{proof}

\begin{proposition}
\label{propconca}Let $S_{\infty}\in\limfunc{Sp}_{\infty}(Z,\sigma)$ have
projection $S=\pi^{\func{Sp}}(S_{\infty})$ such that $\det(S-I)\neq0$ and $%
SX^{\ast}\cap X^{\ast}=0$. Then%
\begin{equation}
\nu(S_{\infty})=\frac{1}{2}(\mu_{X^{\ast}}(S_{\infty})+\limfunc{sign}W_{S})
\label{muw}
\end{equation}
where $W_{S}$ is the symmetric matrix defined by 
\begin{equation*}
W_{S}=DB^{-1}+AB^{-1}-B^{-1}-(B^{T})^{-1}\text{ \ if }S=%
\begin{bmatrix}
A & B \\ 
C & D%
\end{bmatrix}
.
\end{equation*}
\end{proposition}

\begin{proof}
We will divide the proof in three steps. \textit{Step 1.} Let $L\in \func{Lag%
}^{\ominus }(4n,\mathbb{R})$. Using successively formulae (\ref{muis}) and (%
\ref{mule}) we have%
\begin{equation*}
\nu (S_{\infty })=\frac{1}{2}(\mu _{L}^{\ominus }(S_{\infty }^{\ominus
})+\tau ^{\ominus }(S^{\ominus }\Delta ,\Delta ,L)-\tau ^{\ominus
}(S^{\ominus }\Delta ,S^{\ominus }L,L))\text{.}
\end{equation*}%
Choosing in particular $L=L_{0}=X^{\ast }\oplus X^{\ast }$ we get 
\begin{align*}
\mu _{L_{0}}^{\ominus }(S_{\infty }^{\ominus })& =\mu ^{\ominus }((I\oplus
S)_{\infty }(X_{\infty }^{\ast }\oplus X_{\infty }^{\ast }),(X_{\infty
}^{\ast }\oplus X_{\infty }^{\ast })) \\
& =\mu (X_{\infty }^{\ast },X_{\infty }^{\ast })-\mu (X_{\infty }^{\ast
},S_{\infty }X_{\infty }^{\ast }) \\
& =-\mu (X_{\infty }^{\ast },S_{\infty }X_{\infty }^{\ast }) \\
& =\mu _{X^{\ast }}(S_{\infty })
\end{align*}%
so that there remains to prove that 
\begin{equation*}
\tau ^{\ominus }(S^{\ominus }\Delta ,\Delta ,L_{0})-\tau ^{\ominus
}(S^{\ominus }\Delta ,S^{\ominus }L_{0},L_{0})=-\limfunc{sign}W_{S}\text{.}
\end{equation*}%
\textit{Step 2.} We are going to show that%
\begin{equation*}
\tau ^{\ominus }(S^{\ominus }\Delta ,S^{\ominus }L_{0},L_{0})=0\text{;}
\end{equation*}%
in view of the symplectic invariance (K.2) and the antisymmetry (K.1) of $%
\tau ^{\ominus }$ this is equivalent to%
\begin{equation}
\tau ^{\ominus }(L_{0},\Delta ,L_{0},(S^{\ominus })^{-1}L_{0})=0\text{.}
\label{cucu}
\end{equation}%
We have 
\begin{equation*}
\Delta \cap L_{0}=\{(0,p;0,p):p\in \mathbb{R}^{n}\}
\end{equation*}%
and $(S^{\ominus })^{-1}L_{0}\cap L_{0}$ consists of all $(0,p^{\prime
},S^{-1}(0,p^{\prime \prime }))$ with $S^{-1}(0,p^{\prime \prime
})=(0,p^{\prime })$; since $S$ (and hence $S^{-1}$) is free we must have $%
p^{\prime }=p^{\prime \prime }=0$ so that%
\begin{equation*}
(S^{\ominus })^{-1}L_{0}\cap L_{0}=\{(0,p;0,0):p\in \mathbb{R}^{n}\}\text{.}
\end{equation*}%
It follows that we have%
\begin{equation*}
L_{0}=\Delta \cap L_{0}+(S^{\ominus })^{-1}L_{0}\cap L_{0}
\end{equation*}%
hence (\ref{cucu}) in view of property (K.7) of $\tau $. \textit{Step 3}.
Let us finally prove that.%
\begin{equation*}
\tau ^{\ominus }(S^{\ominus }\Delta ,\Delta ,L_{0})=-\limfunc{sign}W_{S}%
\text{;}
\end{equation*}%
this will complete the proof of the proposition. The condition $\det
(S-I)\neq 0$ is equivalent to $S^{\ominus }\Delta \cap \Delta =0$ hence,
using property (K.6) of $\tau $,%
\begin{equation*}
\tau ^{\ominus }(S^{\ominus }\Delta ,\Delta ,L_{0})=-\tau ^{\ominus
}(S^{\ominus }\Delta ,L_{0},\Delta )
\end{equation*}%
is the signature of the quadratic form $Q$ on $L_{0}$ defined by 
\begin{equation*}
Q(0,p,0,p^{\prime })=-\sigma ^{\ominus }(P_{\Delta }(0,p,0,p^{\prime
});0,p,0,p^{\prime })
\end{equation*}%
where 
\begin{equation*}
P_{\Delta }=%
\begin{bmatrix}
(S-I)^{-1} & -(S-I)^{-1}\smallskip \\ 
S(S-I)^{-1} & -S(S-I)^{-1}%
\end{bmatrix}%
\end{equation*}%
is the projection on $S^{\ominus }\Delta $ along $\Delta $ in $Z\oplus Z$.
It follows that the quadratic form $Q$ is given by%
\begin{equation*}
Q(0,p,0,p^{\prime })=-\sigma ^{\ominus }((I-S)^{-1}(0,p^{\prime \prime
}),S(I-S)^{-1}(0,p^{\prime \prime });0,p,0,p^{\prime })
\end{equation*}%
where we have set $p^{\prime \prime }=p-p^{\prime }$; by definition of $%
\sigma ^{\ominus }$ this is%
\begin{multline*}
Q(0,p,0,p^{\prime })= \\
-\sigma ((I-S)^{-1}(0,p^{\prime \prime }),(0,p))+\sigma
(S(I-S)^{-1}(0,p^{\prime \prime }),(0,p^{\prime }))\text{. }
\end{multline*}%
Let now $M_{S}$ be the symplectic Cayley transform (\ref{cayley}) of $S$; we
have%
\begin{equation*}
(I-S)^{-1}=JM_{S}+\tfrac{1}{2}I\text{ \ , \ }S(I-S)^{-1}=JM_{S}-\tfrac{1}{2}I
\end{equation*}%
and hence%
\begin{align*}
Q(0,p,0,p^{\prime })& =-\sigma ((JM_{S}+\tfrac{1}{2}I)(0,p^{\prime \prime
}),(0,p))+\sigma ((JM_{S}-\tfrac{1}{2}I)(0,p^{\prime \prime }),(0,p^{\prime
})) \\
& =-\sigma (JM_{S}(0,p^{\prime \prime }),(0,p))+\sigma (JM_{S}(0,p^{\prime
\prime }),(0,p^{\prime })) \\
& =\sigma (JM_{S}(0,p^{\prime \prime }),(0,p^{\prime \prime })) \\
& =-\left\langle M_{S}(0,p^{\prime \prime }),(0,p^{\prime \prime
})\right\rangle \text{.}
\end{align*}%
Let us calculate explicitly $M_{S}$. Writing $S$ in usual block-form we have%
\begin{equation*}
S-I=%
\begin{bmatrix}
0 & B\smallskip \\ 
I & D-I%
\end{bmatrix}%
\begin{bmatrix}
C-(D-I)B^{-1}(A-I) & 0\smallskip \\ 
B^{-1}(A-I) & I%
\end{bmatrix}%
\end{equation*}%
that is%
\begin{equation*}
S-I=%
\begin{bmatrix}
0 & B \\ 
I & D-I%
\end{bmatrix}%
\begin{bmatrix}
W_{S} & 0 \\ 
B^{-1}(A-I) & I%
\end{bmatrix}%
\end{equation*}%
where we have used the identity 
\begin{equation*}
C-(D-I)B^{-1}(A-I))=B^{-1}A+DB^{-1}-B^{-1}-(B^{T})^{-1}
\end{equation*}%
which follows from the relation $C-DB^{-1}A=-(B^{T})^{-1}$ (the latter is a
rephrasing of the equalities $D^{T}A-B^{T}C=I$ and $D^{T}B=B^{T}D$,
consequences of the fact that $S^{T}JS=S^{T}JS$ since $S\in \limfunc{Sp}%
(Z,\sigma )$). It follows that%
\begin{align*}
(S-I)^{-1}& =%
\begin{bmatrix}
W_{S}^{-1} & 0\smallskip \\ 
B^{-1}(I-A)W_{S}^{-1} & I%
\end{bmatrix}%
\begin{bmatrix}
(I-D)B^{-1} & I\smallskip \\ 
B^{-1} & 0%
\end{bmatrix}%
\medskip \\
& =%
\begin{bmatrix}
W_{S}^{-1}(I-D)B^{-1} & W_{S}^{-1}\smallskip \\ 
B^{-1}(I-A)W_{S}^{-1}(I-D)B^{-1}+B^{-1} & B^{-1}(I-A)W_{S}^{-1}%
\end{bmatrix}%
\end{align*}%
and hence%
\begin{equation*}
M_{S}=%
\begin{bmatrix}
B^{-1}(I-A)W_{S}^{-1}(I-D)B^{-1}+B^{-1} & \frac{1}{2}I+B^{-1}(I-A)W_{S}^{-1}%
\smallskip \\ 
-\frac{1}{2}I-W_{S}^{-1}(I-D)B^{-1} & -W_{S}^{-1}%
\end{bmatrix}%
\end{equation*}%
from which follows that%
\begin{equation*}
Q(0,p,0,p^{\prime })=\left\langle W_{S}^{-1}p^{\prime \prime },p^{\prime
\prime }\right\rangle =\left\langle W_{S}^{-1}(p-p^{\prime }),(p-p^{\prime
})\right\rangle \text{.}
\end{equation*}%
The matrix of the quadratic form $Q$ is thus%
\begin{equation*}
2%
\begin{bmatrix}
W_{S}^{-1} & -W_{S}^{-1}\smallskip \\ 
-W_{S}^{-1} & W_{S}^{-1}%
\end{bmatrix}%
\end{equation*}%
and this matrix has signature $\limfunc{sign}(W_{S}^{-1})=\limfunc{sign}%
W_{S} $, concluding the proof.\bigskip
\end{proof}

\section{The Metaplectic Group}

We denote by $\limfunc{Mp}(Z,\sigma)$ the unitary representation in $%
L^{2}(X) $ of the two-sheeted covering group $\limfunc{Sp}_{2}(Z,\sigma)$ of 
$\limfunc{Sp}(Z,\sigma)$. That group, called the \textit{metaplectic group}
in the literature \cite{Folland,Leray,Wallach}, is generated by the
operators $\widehat{S}_{W,m}$ defined by%
\begin{equation*}
\widehat{S}_{W,m}f(x)=\left( \tfrac{1}{2\pi i}\right) ^{n/2}\Delta
(W)\int_{X}e^{-iW(x,x^{\prime})}f(x^{\prime})dx^{\prime}
\end{equation*}
where 
\begin{equation}
W(x,x^{\prime})=\frac{1}{2}\left\langle Px,x\right\rangle -\left\langle
Lx,x^{\prime}\right\rangle +\frac{1}{2}\left\langle Qx^{\prime},x^{\prime
}\right\rangle  \label{plq}
\end{equation}
with $P=P^{T}$, $Q=Q^{T}$, $\det L\neq0$, and 
\begin{equation*}
\Delta(W)=i^{m}\sqrt{|\det L|}\text{ \ , \ }m\pi=\arg\det L
\end{equation*}
(note that the integer $m$ is only defined modulo $4$). The projection $\pi^{%
\limfunc{Mp}}:\limfunc{Mp}(Z,\sigma)\longrightarrow \limfunc{Sp}(Z,\sigma)$
is determined by the action on the generators $\widehat{S}_{W,m}$ which is
given by $S_{W}=\pi^{\limfunc{Mp}}(\widehat{S}_{W,m})$ where $S_{W}$ is the
free symplectic matrix generated by $W$.

Every $\widehat{S}\in\limfunc{Mp}(Z,\sigma)$ can be written (in infinitely
many ways) as a product $\widehat{S}=\widehat{S}_{W,m}\widehat {S}%
_{W^{\prime},m^{\prime}}$ (see \cite{Leray,Wiley} for a proof); if 
\begin{equation*}
\widehat{S}_{W,m}\widehat{S}_{W^{\prime},m^{\prime}}=\widehat{S}%
_{W^{\prime\prime},m^{\prime\prime}}\widehat{S}_{W^{\prime\prime\prime
},m^{\prime\prime\prime}}
\end{equation*}
then we have 
\begin{align*}
m+m^{\prime}-\limfunc{Inert}(P^{\prime}+Q) & \equiv m^{\prime\prime
}+m^{\prime\prime\prime}-\limfunc{Inert}(P^{\prime\prime\prime
}+Q^{\prime\prime})\text{ \ }\func{mod}4 \\
\limfunc{rank}(P^{\prime}+Q) & \equiv\limfunc{rank}(P^{\prime\prime%
\prime}+Q^{\prime\prime})\text{ \ }\func{mod}4\text{.}
\end{align*}
In \cite{AIF} we have shown that if $\widehat{S}=\widehat{S}_{W,m}\widehat {S%
}_{W^{\prime},m^{\prime}}$ is the projection on $\limfunc{Mp}(Z,\sigma)$ of $%
S_{\infty}\in\limfunc{Sp}_{\infty}(Z,\sigma)$ then 
\begin{align*}
m_{X^{\ast}}(S_{\infty}) & \equiv m+m^{\prime}-\limfunc{Inert}(P^{\prime}+Q)%
\text{ \ }\func{mod}4 \\
& \equiv m+m^{\prime}-\limfunc{Inert}(X^{\ast},S_{W}X^{\ast},S_{W}S_{W^{%
\prime}}X^{\ast})\text{ \ }\func{mod}4\text{;}
\end{align*}
it follows that the class of $m_{X^{\ast}}(S_{\infty})$ modulo $4$ only
depends on the projection $\widehat{S}$; denoting that class by $\widehat {m}%
(\widehat{S})$ the function $\widehat{m}:\limfunc{Mp}(Z,\sigma
)\longrightarrow\mathbb{Z}/4\mathbb{Z}$ is called \textquotedblleft Maslov
index on $\limfunc{Mp}(Z,\sigma)$\textquotedblright. One proves \cite%
{AIF,Wiley} that $\widehat{m}(\widehat{S}_{W,m})=\widehat{m}$ and that%
\begin{equation*}
\widehat{m}(\widehat{S}\widehat{S^{\prime}})=\widehat{m}(\widehat{S})+%
\widehat{m}(\widehat{S^{\prime}})+\widehat{\limfunc{Inert}}(X^{\ast
},SX^{\ast},SS^{\prime}X^{\ast})
\end{equation*}
for all $\widehat{S},\widehat{S^{\prime}}\in\limfunc{Mp}(Z,\sigma)$.

The operators $\widehat{S}_{W,m}$ generate $\limfunc{Mp}(Z,\sigma)$; so do
the operators $\widehat{V}_{P}$, $\widehat{M}_{L,m}$, and $\widehat{J}$
defined by%
\begin{equation*}
\widehat{V}_{P}f(x)=e^{-\frac{i}{2}\left\langle Px,x\right\rangle }f(x)\text{
\ , \ }\widehat{M}_{L,m}f(x)=i^{m}\sqrt{|\det L|}f(Lx)
\end{equation*}
when $P=P^{T}$ and $\det L\neq0$, and%
\begin{equation*}
\widehat{J}f(x)=\left( \tfrac{1}{2\pi i}\right)
^{n/2}\int_{X}e^{-i\left\langle x,x^{\prime}\right\rangle
}f(x^{\prime})dx^{\prime}\text{.}
\end{equation*}
Notice that if $W$ is given by (\ref{plq}) then%
\begin{equation}
\widehat{S}_{W,m}=\widehat{V}_{-P}\widehat{M}_{L,m}\widehat{J}\widehat{V}%
_{-Q}\text{.}  \label{swplq}
\end{equation}

\subsection{A class of unitary operators on $L^{2}(X)$}

We are going to construct a class of Weyl operators $\widehat{R}_{\nu}(S)$
parametrized by $(S,\nu)\in\limfunc{Sp}(Z,\sigma)\times\mathbb{R}$; we will
see that these operators generate a projective representation of $\limfunc{Sp%
}(Z,\sigma)$, containing the metaplectic group $\limfunc{Mp}(Z,\sigma)$
(this last step will be achieved by identifying the parameter $\nu$ with the
index introduced in last section).

Let $S\in\limfunc{Sp}(Z,\sigma)$ be such that $\det(S-I)\neq0$ and define%
\begin{equation}
\widehat{R}_{\nu}(S)=\left( \tfrac{1}{2\pi}\right) ^{n}i^{\nu}\sqrt {%
|\det(S-I)|}\int_{Z}\widehat{T}(Sz)\widehat{T}(-z)dz  \label{alfa1}
\end{equation}
where the integral is interpreted in the sense of Bochner. Taking into
account the relation (\ref{noco2}) we have%
\begin{equation*}
\widehat{T}((S-I)z)=e^{-\frac{i}{2}\sigma(Sz,z)}\widehat{T}(Sz)\widehat {T}%
(-z)
\end{equation*}
so that we can rewrite definition (\ref{alfa1}) as 
\begin{equation}
\widehat{R}_{\nu}(S)=\left( \tfrac{1}{2\pi}\right) ^{n}i^{\nu}\sqrt {%
|\det(S-I)|}\int_{Z}e^{-\frac{i}{2}\sigma(Sz,z)}\widehat{T}((S-I)z)dz\text{.}
\label{alfa2}
\end{equation}
Let us write this formula in Weyl form:

\begin{proposition}
The operator $\widehat{R}_{\nu}(S)$ is given by%
\begin{equation}
\widehat{R}_{\nu}(S)=\left( \frac{1}{2\pi}\right) ^{n}\frac{i^{\nu}}{\sqrt{%
|\det(S-I)|}}\int_{Z}e^{\frac{i}{2}\left\langle M_{S}z,z\right\rangle }%
\widehat{T}(z)dz  \label{alf0}
\end{equation}
where $M_{S}$ is the symplectic Cayley transform of $S$.
\end{proposition}

\begin{proof}
In view of (\ref{cayleybis}) and the antisymmetry of $J$ we have 
\begin{equation*}
\left\langle M_{S}z,z\right\rangle =\left\langle J(S-I)^{-1}z,z\right\rangle
=\sigma((S-I)^{-1}z,z)\text{.}
\end{equation*}
Performing the change of variables $z\longmapsto(S-I)^{-1}z$ we can rewrite
the integral in the right-hand side of (\ref{alfa2}) as%
\begin{align*}
\int_{Z}e^{-\frac{i}{2}\sigma(Sz,z)}\widehat{T}((S-I)z)dz & =\int _{Z}e^{%
\frac{i}{2}\sigma(z,(S-I)z)}\widehat{T}((S-I)z)dz \\
& =\int_{Z}e^{\frac{i}{2}\left\langle M_{S}z,z\right\rangle }\widehat{T}(z)dz
\end{align*}
hence the result.
\end{proof}

Formula (\ref{alf0}) defines a Weyl operator with twisted symbol%
\begin{equation}
a_{\sigma }(z)=\frac{i^{\nu }}{\sqrt{|\det (S-I)|}}e^{\frac{i}{2}%
\left\langle M_{S}z,z\right\rangle }\text{.}  \label{gautwisym}
\end{equation}%
If in addition that $\det (S+I)\neq 0$ we easily deduce from this formula
the usual Weyl symbol $a$. In fact, $a=\mathcal{F}_{\sigma }a_{\sigma }$
that is%
\begin{equation*}
a(z)=\left( \frac{1}{2\pi }\right) ^{n}\frac{i^{\nu }}{\sqrt{|\det (S-I)|}}%
\int_{Z}e^{-i\sigma (z,z^{\prime })}e^{\frac{i}{2}\left\langle
M_{S}z^{\prime },z^{\prime }\right\rangle }dz^{\prime }\text{;}
\end{equation*}%
applying the Fresnel formula (\ref{fres}) with $m=2n$ we then get%
\begin{equation*}
a(z)=\frac{i^{\nu +\frac{1}{2}\limfunc{sign}M_{S}}}{\sqrt{|\det (S-I)|}}%
|\det M_{S}|^{-1/2}e^{\frac{i}{2}\left\langle JM_{S}^{-1}Jz,z\right\rangle }%
\text{.}
\end{equation*}%
Since by definition of $M_{S}$ 
\begin{equation*}
\det M_{S}=2^{-n}\det (S+I)\det (S-I)
\end{equation*}%
we can rewrite the formula above as%
\begin{equation}
a(z)=2^{n/2}\frac{i^{\nu +\frac{1}{2}\limfunc{sign}M_{S}}}{\sqrt{|\det (S+I)|%
}}e^{\frac{i}{2}\left\langle JM_{S}^{-1}Jz,z\right\rangle }\text{.}
\label{gausym}
\end{equation}%
(Behold: this formula is only valid when $S$ has not $-1$ as eigenvalue.)
\medskip

Let us begin by studying composition and inversion for the operators $%
\widehat{R}_{\nu}(S)$. This will allow us to establish that the operators $%
\widehat{R}_{\nu}(S)$ are unitary.

\begin{proposition}
\label{theors}Let $S$ and $S^{\prime}$ in $\limfunc{Sp}(Z,\sigma)$ be such
that $\det(S-I)\neq0$, $\det(S^{\prime}-I)\neq0$. (i) If $\det
(SS^{\prime}-I)\neq0$ then%
\begin{equation}
\widehat{R}_{\nu}(S)\widehat{R}_{\nu}(S^{\prime})=\widehat{R}_{\nu+\nu
^{\prime}+\frac{1}{2}\limfunc{sign}M}(SS^{\prime})\text{.}  \label{ernunu}
\end{equation}
(ii) The operator $\widehat{R}_{\nu}(S)$ is invertible and its inverse is%
\begin{equation}
\widehat{R}_{\nu}(S)^{-1}=\widehat{R}_{-\nu}(S^{-1})\text{.}  \label{erinv}
\end{equation}
\end{proposition}

\begin{proof}
\textit{(i)} The twisted symbols of $\widehat{R}_{\nu }(S)$ and $\widehat{R}%
_{\nu }(S^{\prime })$ are, respectively,%
\begin{align*}
a_{\sigma }(z)& =\frac{i^{\nu }}{\sqrt{|\det (S-I)|}}e^{\frac{i}{2}%
\left\langle M_{S}z,z\right\rangle } \\
b_{\sigma }(z)& =\frac{i^{\nu }}{\sqrt{|\det (S^{\prime }-I)|}}e^{\frac{i}{2}%
\left\langle M_{S^{\prime }}z,z\right\rangle }\text{.}
\end{align*}%
The twisted symbol $c_{\sigma }$ of the compose $\widehat{R}_{\nu }(S)%
\widehat{R}_{\nu }(S^{\prime })$ is given by 
\begin{equation*}
c_{\sigma }(z)=\left( \tfrac{1}{2\pi }\right) ^{n}\int_{Z}e^{\frac{i}{2}%
\sigma (z,z^{\prime })}a_{\sigma }(z-z^{\prime })b_{\sigma }(z^{\prime
})dz^{\prime }
\end{equation*}%
that is%
\begin{equation*}
c_{\sigma }(z)=K\int_{Z}e^{\frac{i}{2}(\sigma (z,z^{\prime })+\Phi
(z,z^{\prime }))}dz^{\prime }
\end{equation*}%
where the constant in front of the integral is 
\begin{equation*}
K=\left( \frac{1}{2\pi }\right) ^{n}\frac{i^{\nu +\nu ^{\prime }}}{\sqrt{%
|\det (S-I)(S^{\prime }-I)|}}
\end{equation*}%
and the phase $\Phi (z,z^{\prime })$ is given by 
\begin{equation*}
\Phi (z,z^{\prime })=\left\langle M_{S}(z-z^{\prime }),z-z^{\prime
}\right\rangle +\left\langle M_{S^{\prime }}z^{\prime },z^{\prime
}\right\rangle
\end{equation*}%
that is%
\begin{equation*}
\Phi (z,z^{\prime })=\left\langle M_{S}z,z\right\rangle -2\left\langle
M_{S}z,z^{\prime }\right\rangle +\left\langle (M_{S}+M_{S^{\prime
}})z^{\prime },z^{\prime }\right\rangle \text{.}
\end{equation*}%
Observing that%
\begin{align*}
\sigma (z,z^{\prime })-2\left\langle M_{S}z,z^{\prime }\right\rangle &
=\left\langle (J-2M_{S})z,z^{\prime }\right\rangle \\
& =-2\left\langle J(S-I)^{-1}z,z^{\prime }\right\rangle
\end{align*}%
we have%
\begin{multline*}
\sigma (z,z^{\prime })+\Phi (z,z^{\prime })=-2\left\langle
J(S-I)^{-1}z,z^{\prime }\right\rangle \\
+\left\langle M_{S}z,z\right\rangle +\left\langle (M_{S}+M_{S^{\prime
}})z^{\prime },z^{\prime }\right\rangle
\end{multline*}%
and hence%
\begin{equation}
c_{\sigma }(z)=Ke^{\frac{i}{2}\left\langle M_{S}z,z\right\rangle
)}\int_{Z}e^{-i\left\langle J(S-I)^{-1}z,z^{\prime }\right\rangle }e^{\frac{i%
}{2}\left\langle (M_{S}+M_{S^{\prime }})z^{\prime },z^{\prime }\right\rangle
}dz^{\prime }\text{.}  \label{bonin}
\end{equation}%
Applying the Fresnel formula (\ref{fres}) with $m=2n$ to the formula above
and replacing $K$ with its value we get%
\begin{equation}
c_{\sigma }(z)=\left( \tfrac{1}{2\pi }\right) ^{n}|\det [(M_{S}+M_{S^{\prime
}})(S-I)(S^{\prime }-I)]|^{-1/2}e^{\frac{i\pi }{4}\limfunc{sign}M}e^{i\Theta
(z)}  \label{cestca}
\end{equation}%
where the phase $\Theta $ is given by 
\begin{align*}
\Theta (z)& =\left\langle M_{S}z,z\right\rangle -\left\langle
(M_{S}+M_{S^{\prime }})^{-1}J(S-I)^{-1}z,J(S-I)^{-1}z\right\rangle \\
& =\left\langle M_{S}+(S^{T}-I)^{-1}J(M_{S}+M_{S^{\prime
}})^{-1}J(S-I)^{-1}z,z\right\rangle
\end{align*}%
that is $\Theta (z)=M_{SS^{\prime }}$ in view of part \textit{(ii)} of Lemma %
\ref{lemcarton}. Noting that by definition (\ref{cayleybis}) of the
symplectic Cayley transform we have%
\begin{equation*}
M_{S}+M_{S^{\prime }}=J(I+(S-I)^{-1}+(S^{\prime }-I)^{-1})
\end{equation*}%
it follows, using property (\ref{mss}) of the symplectic Cayley transform,
that 
\begin{align*}
\det [(M_{S}+M_{S^{\prime }})(S-I)(S^{\prime }-I)]& =\det
[(S-I)(M_{S}+M_{S^{\prime }})(S^{\prime }-I)] \\
& =\det [(S-I)(M_{S}+M_{S^{\prime }})(S^{\prime }-I)] \\
& =|\det (SS^{\prime }-I)|
\end{align*}%
which concludes the proof of the first part of proposition. Proof of \textit{%
(ii)}. Since $\det (S-I)\neq 0$ we also have $\det (S^{-1}-I)\neq 0$.
Formula (\ref{bonin}) in the proof of part \textit{(i)} shows that the
symbol of $\widehat{C}=\widehat{R}_{\nu }(S)\widehat{R}_{-\nu }(S^{-1})$ is 
\begin{equation*}
c_{\sigma }(z)=Ke^{\frac{i}{2}\left\langle M_{S}z,z\right\rangle
)}\int_{Z}e^{-i\left\langle J(S-I)^{-1}z,z^{\prime }\right\rangle }e^{\frac{i%
}{2}\left\langle (M_{S}+M_{S^{-1}})z^{\prime },z^{\prime }\right\rangle
}dz^{\prime }
\end{equation*}%
where the constant $K$ is this time%
\begin{align*}
K& =\left( \frac{1}{2\pi }\right) ^{n}\frac{1}{\sqrt{|\det (S-I)(S^{-1}-I)|}}
\\
& =\left( \frac{1}{2\pi }\right) ^{n}\frac{1}{|\det (S-I)|}
\end{align*}%
since $\det (S^{-1}-I)=\det (I-S)$. Using again Lemma \ref{lemcarton} we
have $M_{S}+M_{S^{-1}}=0$ hence, setting $z^{\prime \prime
}=(S^{T}-I)^{-1}Jz^{\prime }$,%
\begin{align*}
c_{\sigma }(z)& =\left( \frac{1}{2\pi }\right) ^{n}\frac{e^{\frac{i}{2}%
\left\langle M_{S}z,z\right\rangle }}{|\det (S-I)|}\int_{Z}e^{-i\left\langle
J(S-I)^{-1}z,z^{\prime }\right\rangle }dz^{\prime } \\
& =\left( \frac{1}{2\pi }\right) ^{n}e^{\frac{i}{2}\left\langle
M_{S}z,z\right\rangle )}\int_{Z}e^{i\left\langle z,z^{\prime \prime
}\right\rangle }dz^{\prime \prime } \\
& =(2\pi )^{n}\delta (z)
\end{align*}%
and $\widehat{C}$ is thus the identity operator.\bigskip
\end{proof}

The composition formula above allows us to prove that the operators $%
\widehat{R}_{\nu}(S)$ are unitary:

\begin{corollary}
Let $S\in\limfunc{Sp}(Z,\sigma)$ be such that $\det(S-I)\neq0$. The
operators $\widehat{R}_{\nu}(S)$ are unitary: $\widehat{R}_{\nu}(S)^{\ast}=$ 
$\widehat{R}_{\nu}(S)^{-1}$.
\end{corollary}

\begin{proof}
The symbol of the adjoint of a Weyl operator is the complex conjugate of the
symbol of that operator. Since the twisted and Weyl symbol are symplectic
Fourier transforms of each other the symbol $a$ of $\widehat{R}_{\nu}(S)$ is
thus given by%
\begin{equation*}
\left( 2\pi\right) ^{n}a(z)=\frac{i^{\nu}}{\sqrt{|\det(S-I)|}}\int
_{Z}e^{-i\sigma(z,z^{\prime})}e^{\frac{i}{2}\left\langle M_{S}z^{\prime
},z^{\prime}\right\rangle }dz^{\prime}\text{.}
\end{equation*}
We have%
\begin{equation*}
\left( 2\pi\right) ^{n}\overline{a(z)}=\frac{i^{-\nu}}{\sqrt{|\det(S-I)|}}%
\int_{Z}e^{i\sigma(z,z^{\prime})}e^{-\frac{i}{2}\left\langle M_{S}z^{\prime
},z^{\prime}\right\rangle }dz^{\prime}\text{.}
\end{equation*}
Since $M_{S^{-1}}=-M_{S}$ and $|\det(S-I)|=|\det(S^{-1}-I)|$ we have 
\begin{align*}
\left( 2\pi\right) ^{n}\overline{a(z)} & =\frac{i^{-\nu}}{\sqrt {%
|\det(S^{-1}-I)|}}\int_{Z}e^{-i\sigma(z,z^{\prime})}e^{\frac{i}{2}%
\left\langle M_{S^{-1}}z^{\prime},z^{\prime}\right\rangle }dz^{\prime} \\
& =\frac{i^{-\nu}}{\sqrt{|\det(S^{-1}-I)|}}\int_{Z}e^{i\sigma(z,z^{%
\prime})}e^{\frac{i}{2}\left\langle
M_{S^{-1}}z^{\prime},z^{\prime}\right\rangle }dz^{\prime}
\end{align*}
hence $\overline{a(z)}$ is the symbol of $\widehat{R}_{\nu}(S)^{-1}$ and
this concludes the proof.
\end{proof}

\subsection{Relation with $\limfunc{Mp}(Z,\protect\sigma)$\label{susecmpcz}}

Let $S_{\infty}\in\limfunc{Sp}(Z,\sigma)$ have projection $\pi ^{\limfunc{Sp}%
}(S_{\infty})=S$. Proposition \ref{theors} and its Corollary will allow us
to prove that if we choose $\nu=\nu(S_{\infty})$ in $\widehat{R}_{\nu}(S)$
then this operator is in the metaplectic group $\limfunc{Mp}(Z,\sigma)$. The
proof of this property will however require some work. Let us begin by
giving a definition: Let $\widehat{S}\in\limfunc{Mp}(Z,\sigma)$ have
projection $S\in\limfunc{Sp}(Z,\sigma)$ such that $\det(S-I)\neq0$ and
choose $S_{\infty}\in \limfunc{Sp}_{\infty}(Z,\sigma)$ covering $\widehat{S}$%
. We define%
\begin{equation}
\widehat{\nu}(\widehat{S})\equiv\nu(S_{\infty})\text{ \ }\func{mod}4\text{.}
\label{defczmp}
\end{equation}

The index $\widehat{\nu}$ is well-defined: assume in fact that $S_{\infty
}^{\prime}$ is a second element of $\limfunc{Sp}_{\infty}(Z,\sigma)$
covering $\widehat{S}$; we have $S_{\infty}^{\prime}=\alpha^{r}S_{\infty}$
for some $r\in\mathbb{Z}$ ($\alpha$ the generator of $\pi_{1}[\limfunc{Sp}%
(Z,\sigma)]$); since $\limfunc{Mp}(Z,\sigma)$ is a double covering of $%
\limfunc{Sp}(Z,\sigma)$ the integer $r$ must be even. Recalling that 
\begin{equation*}
\nu(\alpha^{r}S_{\infty})=\nu(S_{\infty})+2r
\end{equation*}
the left-hand side of (\ref{defczmp}) only depends on $\widehat{S}$ and not
on the element of $\limfunc{Sp}_{\infty}(Z,\sigma)$ covering it.

Let $S$ and $S^{\prime }$ in $\limfunc{Sp}(Z,\sigma )$ be such that $\det
(S-I)\neq 0$. Let $\widehat{S}$ and $\widehat{S}^{\prime }$ in $\limfunc{Mp}%
(Z,\sigma )$ have projections $S$ and $S^{\prime }$: $\pi ^{\limfunc{Mp}}(%
\widehat{S})=S$ and $\pi ^{\limfunc{Mp}}(\widehat{S}^{\prime })=S^{\prime }$
(there are two possible choices in each case). We have (product property of
the Conley--Zehnder index) 
\begin{equation*}
\nu (S_{\infty }S_{\infty }^{\prime })=\nu (S_{\infty })+\nu (S_{\infty
}^{\prime })+\frac{1}{2}\limfunc{sign}(M_{S}+M_{S^{\prime }})
\end{equation*}%
hence, taking classes modulo $4$, 
\begin{equation*}
\widehat{\nu }(\widehat{S}\widehat{S}^{\prime })=\widehat{\nu }(\widehat{S})+%
\widehat{\nu }(\widehat{S^{\prime }})+\frac{1}{2}\widehat{\limfunc{sign}}%
(M_{S}+M_{S^{\prime }})\text{.}
\end{equation*}%
Choosing $\nu =\nu (\widehat{S})$, $\nu ^{\prime }=\nu (\widehat{S}^{\prime
})$ formula (\ref{ernunu}) becomes%
\begin{equation}
\widehat{R}_{\nu (\widehat{S})}(S)\widehat{R}_{\nu (\widehat{S}^{\prime
})}(S^{\prime })=\widehat{R}_{\nu (\widehat{S}\widehat{S}^{\prime
})}(SS^{\prime })  \label{ernunubis}
\end{equation}%
which suggests that the operators $\widehat{R}_{\nu (\widehat{S})}(S)$
generate a true (two-sheeted) unitary representation of the symplectic
group, that is the metaplectic group. Formula (\ref{ernunubis}) is however
not sufficient to prove this, because the $\widehat{R}_{\nu (\widehat{S}%
)}(S) $ have only been defined for $\det (S-I)\neq 0$. We are going to show
that these operator generate a group, and that this group is indeed the
metaplectic group $\limfunc{Mp}(Z,\sigma )$.

Recall that if $W$ is a quadratic form (\ref{plq}) we denoted by $W_{S}$ the
Hessian matrix of the function $x\longmapsto W(x,x)$:%
\begin{equation}
W_{S}=P+Q-L-L^{T}  \label{wxx}
\end{equation}
that is%
\begin{equation}
W_{S}=DB^{-1}+B^{-1}A-B^{-1}-(B^{T})^{-1}  \label{wxxx}
\end{equation}
where $S=%
\begin{bmatrix}
A & B \\ 
C & D%
\end{bmatrix}
$ is the free symplectic matrix generated by $W$. Also recall (Lemma \ref%
{lemma1}) that 
\begin{align}
\det(S-I) & =(-1)^{n}\det B\det(B^{-1}A+DB^{-1}-B^{-1}-(B^{T})^{-1})
\label{bofor} \\
& =(-1)^{n}\det L^{-1}\det(P+Q-L-L^{T})\text{.}  \notag
\end{align}

We begin by proving that $\widehat{R}_{\nu}(S_{W})$ can be identified with $%
\widehat{S}_{W,m}$ if $\nu$ is chosen in a suitable way:

\begin{proposition}
\label{proplett}Let $\widehat{S}_{W,m}\in\limfunc{Mp}(Z,\sigma)$ be one of
the two operators with projection $S=S_{W}$. (i) We have $\widehat{R}_{\nu
}(S_{W})=\widehat{S}_{W,m}$ provided that%
\begin{equation}
\nu\equiv\nu(\widehat{S})\text{ \ }\func{mod}4\text{;}  \label{Maslov1}
\end{equation}
(ii) When this is the case we have%
\begin{equation}
\arg\det(S-I)\equiv(\nu(\widehat{S})-n)\pi\text{ \ }\func{mod}2\pi\text{.}
\label{Maslov2}
\end{equation}
\end{proposition}

\begin{proof}
Proof of \textit{(i)}. Let $\delta\in\mathcal{S}^{\prime}(\mathbb{R}^{n})$
be the Dirac distribution centered at $x=0$; setting%
\begin{equation*}
C_{W,\nu}=\left( \frac{1}{2\pi}\right) ^{n}\frac{i^{\nu}}{\sqrt{|\det (S-I)|}%
}
\end{equation*}
we have, by definition of $\widehat{R}_{\nu}(S)$, 
\begin{align*}
\widehat{R}_{\nu}(S)\delta(x) & =C_{W,\nu}\int_{Z}e^{\frac{i}{2}\left\langle
M_{S}z_{0},z_{0}\right\rangle }e^{i(\left\langle p_{0},x\right\rangle -\frac{%
1}{2}\left\langle p_{0},x_{0}\right\rangle )}\delta(x-x_{0})dz_{0} \\
& =C_{W,\nu}\int_{Z}e^{\frac{i}{2}\left\langle
M_{S}(x,p_{0}),(x,p_{0})\right\rangle }e^{\frac{i}{2}\left\langle
p,x\right\rangle }\delta (x-x_{0})dz_{0}
\end{align*}
hence, setting $x=0$,%
\begin{equation*}
\widehat{R}_{\nu}(S)\delta(0)=C_{W,\nu}\int_{Z}e^{\frac{i}{2}\left\langle
M_{S}(0,p_{0}),(0,p_{0})\right\rangle }\delta(-x_{0})dz_{0}
\end{equation*}
that is, since $\int_{X}\delta(-x_{0})dx_{0}=1$,%
\begin{equation}
\widehat{R}_{\nu}(S)\delta(0)=\left( \frac{1}{2\pi}\right) ^{n}\frac{i^{\nu }%
}{\sqrt{|\det(S-I)|}}\int_{Z}e^{\frac{i}{2}\left\langle
M_{S}(0,p_{0}),(0,p_{0})\right\rangle }dp_{0}\text{.}  \label{sdo}
\end{equation}
Let us next calculate the scalar product 
\begin{equation*}
\left\langle M_{S}(0,p_{0}),(0,p_{0})\right\rangle
=\sigma((S-I)^{-1}0,p_{0}),(0,p_{0}))\text{.}
\end{equation*}
The relation $(x,p)=(S-I)^{-1}(0,p_{0})$ is equivalent to $%
S(x,p)=(x,p+p_{0}) $ that is to%
\begin{equation*}
p+p_{0}=\partial_{x}W(x,x)\text{ \ and \ }p=-\partial_{x^{\prime}}W(x,x)%
\text{.}
\end{equation*}
these relations yield after a few calculations%
\begin{equation*}
x=(P+Q-L-L^{T})^{-1}p_{0}\text{ \ ; \ }p=(L-Q)(P+Q-L-L^{T})^{-1}p_{0}
\end{equation*}
and hence%
\begin{equation}
\left\langle M_{S}(0,p_{0}),(0,p_{0})\right\rangle =-\left\langle
W_{S}^{-1}p_{0},p_{0}\right\rangle  \label{bofor3}
\end{equation}
where $W_{S}$ is the symmetric matrix (\ref{wxx}). Applying the Fresnel
formula (\ref{fres}) to the integral in (\ref{sdo}) we get%
\begin{equation*}
\left( \tfrac{1}{2\pi}\right) ^{n}\int_{X^{\ast}}e^{\frac{i}{2}\left\langle
M_{S}(0,p_{0}),(0,p_{0})\right\rangle }dp_{0}=e^{-\frac{i\pi}{4}\limfunc{sign%
}W_{S}}|\det W_{S}|^{1/2}\text{;}
\end{equation*}
observing that in view of formula (\ref{bofor}) we have 
\begin{equation*}
\frac{1}{\sqrt{|\det(S_{W}-I)|}}=|\det L|^{1/2}|\det W_{S}|^{-1/2}
\end{equation*}
we obtain%
\begin{equation*}
\widehat{R}_{\nu}(S_{W})\delta(0)=\left( \tfrac{1}{2\pi}\right) ^{n}i^{\nu
}e^{-\frac{i\pi}{4}\limfunc{sign}W_{S}}|\det L|^{1/2}\text{.}
\end{equation*}
Now, by definition of $\widehat{S}_{W,m}$,%
\begin{align*}
\widehat{S}_{W,m}\delta(0) & =\left( \tfrac{1}{2\pi i}\right) ^{n}i^{m}\sqrt{%
|\det L|}\int_{X}e^{iW(0,x^{\prime})}\delta(x^{\prime})dx^{\prime } \\
& =\left( \tfrac{1}{2\pi}\right) ^{n}i^{m-n/2}\sqrt{|\det L|}
\end{align*}
and hence%
\begin{equation*}
i^{\nu}e^{-\frac{i\pi}{4}\limfunc{sign}W_{S}}=i^{m-n/2}\text{.}
\end{equation*}
It follows that we have%
\begin{equation*}
\nu-\frac{1}{2}\limfunc{sign}W_{S}\equiv m-\frac{n}{2}\text{ \ }\func{mod}4
\end{equation*}
which is equivalent to formula (\ref{Maslov1}) since $W_{S}$ has rank $n$.
Proof of \textit{(ii)}. In view of formula (\ref{bofor}) we have%
\begin{equation*}
\arg\det(S-I)=n\pi+\arg\det B+\arg\det W_{S}\text{ \ }\func{mod}2\pi\text{.}
\end{equation*}
Taking into account the obvious relations%
\begin{align*}
\arg\det B & \equiv\pi\widehat{m}(\widehat{S})\text{ \ }\func{mod}2\pi \\
\arg\det W_{S} & \equiv\pi\limfunc{Inert}W_{S}\text{ \ }\func{mod}2\pi
\end{align*}
formula (\ref{Maslov2}) follows.\bigskip
\end{proof}

Recall that $\widehat{S}\in\limfunc{Mp}(Z,\sigma)$ can be written (in
infinitely many ways) as a product $\widehat{S}=\widehat{S}_{W,m}\widehat {S}%
_{W^{\prime},m^{\prime}}$. We are going to show that $\widehat{S}_{W,m}$ and 
$\widehat{S}_{W^{\prime},m^{\prime}}$ always can be chosen such that $\det(%
\widehat{S}_{W,m}-I)\neq0$ and $\det(\widehat{S}_{W^{\prime},m^{\prime}}-I)%
\neq0$.

\begin{corollary}
\label{propabove}The operators $\widehat{R}_{\nu}(S_{W})$ generate $\limfunc{%
Mp}(Z,\sigma)$. In fact, every $\widehat{S}\in \limfunc{Mp}(Z,\sigma)$ can
be written as a product 
\begin{equation}
\widehat{S}=\widehat{S}_{W,m}\widehat{S}_{W^{\prime},m^{\prime}}=\widehat {R}%
_{\nu}(S_{W})\widehat{R}_{\nu^{\prime}}(S_{W^{\prime}})  \label{rsw}
\end{equation}
where $\det(S_{W}-I)\neq0$, $\det(S_{W^{\prime}}-I)\neq0$, and $\nu$, $%
\nu^{\prime}$ are given by (\ref{Maslov1}).
\end{corollary}

\begin{proof}
Let $\widehat{S}=\widehat{S}_{W,m}\widehat{S}_{W^{\prime},m^{\prime}}$. In
view of Proposition \ref{proplett} it suffices to show that $W$ and $%
W^{\prime}$ can be chosen so that $S_{W}=\pi^{\limfunc{Mp}}(\widehat {S}%
_{W,m})$ and $S_{W^{\prime}}=\pi^{\limfunc{Mp}}(\widehat {S}_{W^{\prime},m})$
satisfy $\det(S_{W}-I)\neq0$, $\det(S_{W^{\prime}}-I)\neq0$. That the $%
\widehat{R}_{\nu}(S_{W})$ indeed generate $\limfunc{Mp}(Z,\sigma)$ follows
from formula (\ref{rsw}). Let us write $\widehat{S}=\widehat{S}_{W,m}%
\widehat{S}_{W^{\prime},m^{\prime}}$ and apply the factorization (\ref{swplq}%
) to each of the factors:%
\begin{equation}
\widehat{S}=\widehat{V}_{-P}\widehat{M}_{L,m}\widehat{J}\widehat {V}%
_{-(P^{\prime}+Q)}\widehat{M}_{L^{\prime},m^{\prime}}\widehat{J}\widehat {V}%
_{-Q^{\prime}}\text{.}  \label{sprod}
\end{equation}
We claim that $\widehat{S}_{W,m}$ and $\widehat{S}_{W^{\prime},m^{\prime}}$
can be chosen in such a way that $\det(S_{W}-I)\neq0$ and $%
\det(S_{W^{\prime}}-I)\neq0$ that is, 
\begin{equation*}
\det(P+Q-L-L^{T})\neq0\text{ \ and \ }\det(P^{\prime}+Q^{\prime}-L^{\prime
}-L^{\prime T})\neq0\text{;}
\end{equation*}
this will prove the assertion. We first remark that the right hand-side of (%
\ref{sprod}) obviously does not change if we replace $P^{\prime}$ by $%
P^{\prime}+\lambda I$ and $Q$ by $Q-\lambda I$ where $\lambda\in\mathbb{R}$.
Choose now $\lambda$ such that it is not an eigenvalue of $P+Q-L-L^{T}$ and $%
-\lambda$ is not an eigenvalue of $P^{\prime}+Q^{\prime}-L^{\prime}-L^{%
\prime T}$; then 
\begin{align*}
\det(P+Q-\lambda I-L-L^{T}) & \neq0 \\
\det(P^{\prime}+\lambda I+Q^{\prime}-L-L^{T}) & \neq0
\end{align*}
and we have $\widehat{S}=\widehat{S}_{W_{1},m_{1}}\widehat{S}_{W_{1}^{\prime
},m_{1}^{\prime}}$ with%
\begin{align*}
W_{1}(x,x^{\prime}) & =\frac{1}{2}\langle Px,x\rangle-\langle Lx,x^{\prime
}\rangle+\frac{1}{2}\langle(Q-\lambda I)x^{\prime},x^{\prime}\rangle \\
W_{1}^{\prime}(x,x^{\prime}) & =\frac{1}{2}\langle(P^{\prime}+\lambda
I)x,x\rangle-\langle L^{\prime}x,x^{\prime}\rangle+\frac{1}{2}\langle
Q^{\prime}x^{\prime},x^{\prime}\rangle\text{;}
\end{align*}
this concludes the proof.\bigskip
\end{proof}

There remains to prove that every $\widehat{S}\in\limfunc{Mp}(Z,\sigma)$
such that $\det(S-I)\neq0$ can be written in the form $\widehat{R}_{\nu}(S)$:

\begin{proposition}
For every $\widehat{S}\in\limfunc{Mp}(Z,\sigma)$ such that $\det(S-I)\neq0$
we have $\widehat{S}=\widehat{R}_{\nu(\widehat{S})}(S)$ with 
\begin{equation}
\nu(\widehat{S})=\nu+\nu^{\prime}+\tfrac{1}{2}\limfunc{sign}(M+M^{\prime})
\label{srs}
\end{equation}
if $\widehat{S}=\widehat{R}_{\nu}(S_{W})\widehat{R}_{\nu^{\prime}}(S_{W^{%
\prime}})$ and $M=M_{S_{W}}$, $M^{\prime}=M_{S_{W^{\prime}}}$.
\end{proposition}

\begin{proof}
Let us write $\widehat{S}=\widehat{R}_{\nu }(S_{W})\widehat{R}_{\nu ^{\prime
}}(S_{W^{\prime }})$. A straightforward calculation using the composition
formula (\ref{ernunu}) and the Fresnel integral (\ref{fres}) shows that 
\begin{equation}
\widehat{S}=\left( \frac{1}{2\pi }\right) ^{n}\frac{i^{\nu +\nu ^{\prime }+%
\frac{1}{2}sgn(M+M^{\prime })}}{\sqrt{|\det (S_{W}-I)(S_{W^{\prime
}}-I)(M+M^{\prime })|}}\int_{Z}e^{\frac{i}{2}\left\langle Nz,z\right\rangle }%
\widehat{T}(z)dz  \label{ssm}
\end{equation}%
where $N$ is given by 
\begin{equation*}
N=M-(M+\tfrac{1}{2}J)(M+M^{\prime })^{-1}(M-\tfrac{1}{2}J)\text{.}
\end{equation*}%
We claim that%
\begin{equation}
\det (S_{W}-I)(S_{W^{\prime }}-I)(M+M^{\prime })=\det (S-I)  \label{cl1}
\end{equation}%
(hence $M+M^{\prime }$ is indeed invertible), and that%
\begin{equation}
N=\tfrac{1}{2}J(S+I)(S-I)^{-1}=M_{S}\text{.}  \label{cl2}
\end{equation}%
The first of these identities is easy to check by a direct calculation: by
definition of $M$ and $M^{\prime }$ we have, since $\det J=1$,%
\begin{multline*}
\det (S_{W}-I)(S_{W^{\prime }}-I)(M+M^{\prime })= \\
\det (S_{W}-I)(I+(S_{W}-I)^{-1}+(S_{W}-I)^{-1})(S_{W^{\prime }}-I)
\end{multline*}%
that is%
\begin{equation*}
\det (S_{W}-I)(S_{W^{\prime }}-I)(M+M^{\prime })=\det (S_{W}S_{W^{\prime
}}-I)
\end{equation*}%
which is precisely (\ref{cl1}). Formula (\ref{cl2}) is at first sight more
cumbersome; there is however an easy way out: assume that $\widehat{S}=%
\widehat{S}_{W^{\prime \prime },m^{\prime \prime }}$; in view of Lemma \ref%
{lemcarton} we have in this case 
\begin{equation*}
N=\tfrac{1}{2}J(S_{W}S_{W^{\prime }}+I)(S_{W}S_{W^{\prime }}-I)^{-1}
\end{equation*}%
and this algebraic identity then holds for all $S=S_{W}S_{W^{\prime }}$
since the free symplectic matrices are dense in $\limfunc{Sp}(Z,\sigma )$.
Thus,%
\begin{equation*}
\widehat{S}=\left( \frac{1}{2\pi }\right) ^{n}\frac{i^{\nu +\nu ^{\prime }+%
\frac{1}{2}sgn(M+M^{\prime })}}{\sqrt{|\det (S-I)|}}\int_{Z}e^{\frac{i}{2}%
\left\langle M_{S}z,z\right\rangle }\widehat{T}(z)dz
\end{equation*}%
and to conclude the proof there remains to prove that 
\begin{equation*}
\nu (S)\pi =(\nu +\nu ^{\prime }+\tfrac{1}{2}\limfunc{sign}(M+M^{\prime
}))\pi
\end{equation*}%
is effectively one of the two possible choices for $\arg \det (S-I)$. We have%
\begin{multline*}
(\nu +\nu ^{\prime }+\tfrac{1}{2}\limfunc{sign}(M+M^{\prime }))\pi = \\
-\arg \det (S_{W}-I)-\arg \det (S_{W^{\prime }}-I)+\tfrac{1}{2}\pi \limfunc{%
sign}(M+M^{\prime });
\end{multline*}%
we next note that if $R$ is any real invertible $2n\times 2n$ symmetric
matrix with $q$ negative eigenvalues we have $\arg \det R=q\pi $ $\func{mod}%
2\pi $ and $\frac{1}{2}\limfunc{sign}R=2n-q$ and hence%
\begin{equation*}
\arg \det R=\tfrac{1}{2}\pi \limfunc{sign}R\text{ \ }\func{mod}2\pi \text{.}
\end{equation*}%
It follows, taking (\ref{cl1}) into account, that 
\begin{equation*}
(\nu +\nu ^{\prime }+\tfrac{1}{2}\limfunc{sign}(M+M^{\prime }))\pi =\arg
\det (S-I)\text{ \ }\func{mod}2\pi
\end{equation*}%
which concludes the proof.
\end{proof}

\section{Weyl Calculus on Symplectic Space}

Let us now define a class of pseudo-differential operators acting on
functions defined on $(Z,\sigma)$. The passage from the usual Weyl calculus
is made explicit using a family of isometries of $L^{2}(X\mathcal{)}$ onto
closed subspaces of $L^{2}(Z\mathcal{)}$. Using the results of previous
section we will establish that the calculus thus constructed enjoys a
property of metaplectic covariance which makes it into a true generalization
of the usual Weyl calculus.

\subsection{The isometries $U_{\protect\phi}$}

In what follows $\phi\in\mathcal{S}(X)$ is normalized to the unity: $%
||\phi||_{L^{2}(X)}^{2}=1.$ We associate to $\phi$ the integral operator $%
U_{\phi}:L^{2}(X\mathcal{)}\longrightarrow L^{2}(Z\mathcal{)}$ defined by 
\begin{equation}
U_{\phi}f(z)=\left( \tfrac{\pi}{2}\right) ^{n/2}W(f,\overline{\phi})(\tfrac{1%
}{2}z)\text{.}  \label{defwpt}
\end{equation}
where $W(f,\overline{\phi})$ is the Wigner--Moyal transform (\ref{wm}) of
the pair $(f,\overline{\phi}).$ A standard --but by no means mandatory--
choice is to take for $\phi$ the real Gaussian%
\begin{equation}
\phi_{0}(x)=\left( \tfrac{1}{\pi}\right) ^{n/4}e^{-\frac{1}{2}|x|^{2}}\text{;%
}  \label{fizero}
\end{equation}
the corresponding operator $U_{\phi}$ is then (up to an exponential factor)
the \textquotedblleft coherent state representation\textquotedblright\
familiar to quantum physicists.

\begin{proposition}
\label{propun}The transform $U_{\phi}$ has the following properties: (i) $%
U_{\phi}$ is an isometry: the Parseval formula%
\begin{equation}
(U_{\phi}f,U_{\phi}f^{\prime})_{L^{2}(Z)}=(f,f^{\prime})_{L^{2}(X)}
\label{parseval}
\end{equation}
holds for all $f,f^{\prime}\in\mathcal{S}(X)$. In particular $U_{\phi}^{\ast
}U_{\phi}=I$ on\ $L^{2}(X)$. (ii) The range $\mathcal{H}_{\phi}$ of $%
U_{\phi} $ is closed in $L^{2}(Z)$ (and is hence a Hilbert space), and the
operator $P_{\phi}=U_{\phi}U_{\phi}^{\ast}$ is the orthogonal projection in $%
L^{2}(Z)$ onto $\mathcal{H}_{\phi}$. (iii) Let $\widehat{S}\in\limfunc{Mp}%
(Z,\sigma)$, $\pi^{\limfunc{Mp}}(\widehat{S})=S$. We have%
\begin{equation}
U_{\phi}(\widehat{S}f)=(U_{\phi_{\widehat{S}}}f)\circ S^{-1}\text{ \ , \ }%
\phi_{\widehat{S}}=\overline{\widehat{S}^{-1}\overline{\phi}}\text{.}
\label{mcwpt}
\end{equation}
\end{proposition}

\begin{proof}
\textit{(i)} Formula (\ref{parseval}) is an immediate consequence of the
property%
\begin{equation}
(W(f,\phi),W(f^{\prime},\phi^{\prime}))_{L^{2}(Z)}=\left( \tfrac{1}{2\pi }%
\right) ^{n}(f,f^{\prime})_{L^{2}(X)}\overline{(\phi,\phi^{%
\prime})_{L^{2}(X)}}  \label{important}
\end{equation}
of the Wigner--Moyal transform (see \textit{e.g. }Folland \cite{Folland} 
\textit{p}. 56). \textit{(ii)} It is clear that $P_{\phi}^{2}=P_{\phi}$. Let
us show that the range of $P_{\phi}$ is $\mathcal{H}_{\phi}$; the closedness
of $\mathcal{H}_{\phi}$ will follow. Since $U_{\phi}^{\ast}U_{\phi}=I$ on $%
L^{2}(X)$ we have $U_{\phi}^{\ast}U_{\phi}f=f$ for every $f$ in $L^{2}(X)$
and hence the range of $U_{\phi}^{\ast}$ is $L^{2}(X)$. It follows that the
range of $U_{\phi}$ is that of $U_{\phi}U_{\phi}^{\ast}=P_{\phi}$ and is
hence closed. Recalling that the Wigner--Moyal transform is such that 
\begin{equation}
W(\widehat{S}f,\widehat{S}\phi)=W(f,\phi)\circ S^{-1}  \label{metac1}
\end{equation}
for every $\widehat{S}\in\limfunc{Mp}(Z,\sigma)$ with $\pi ^{\limfunc{Mp}}(%
\widehat{S})=S$ we have, using definition (\ref{defwpt}) of $U_{\phi}$,%
\begin{align*}
U_{\phi}(\widehat{S}f) & =\left( \tfrac{\pi}{2}\right) ^{n/2}W(\widehat {S}f,%
\overline{\phi})(\tfrac{1}{2}z) \\
& =\left( \tfrac{\pi}{2}\right) ^{n/2}W(\widehat{S}f,\widehat{S}(\widehat{S}%
^{-1}\overline{\phi}))(\tfrac{1}{2}z) \\
& =\left( \tfrac{\pi}{2}\right) ^{n/2}W(f,\widehat{S}^{-1}\overline{\phi }))(%
\tfrac{1}{2}S^{-1}(z))
\end{align*}
hence (%
\index{mcwpt}).\bigskip
\end{proof}

The observant reader will perhaps remember from the Introduction that the
operator $%
\widehat{T}_{\text{ph}}(z_{0})$ was obtained by formally replacing $z$ in $%
\sigma(z,z_{0})$ by operator $\widehat{z}_{\text{ph}}=(\widehat {x}_{\text{ph%
}},\widehat{p}_{\text{ph}})$ where 
\begin{equation}
\widehat{x}_{\text{ph}}=\tfrac{1}{2}x+i\partial_{p}\text{ \ , \ }\widehat {x}%
_{\text{ph}}=\tfrac{1}{2}p-i\partial_{x}  \label{quantumrule}
\end{equation}
(formula (%
\index{quru}). In addition, for every transform $U_{\phi}$ we have%
\begin{equation}
U_{\phi}(xf)=%
\widehat{x}_{\text{ph}}U_{\phi}(f)\text{ , }U_{\phi}(-i\partial_{x}f)=%
\widehat{x}_{\text{ph}}U_{\phi}(f)\text{ }  \label{uffe}
\end{equation}
for all $f\in\mathcal{S}(X)$; the proof is purely computational and left to
the reader.

One should be aware of the fact that the Hilbert space $\mathcal{H}_{\phi}$
is smaller than $L^{2}(Z)$:

\begin{example}
\label{theorange}Assume that $\phi=\phi_{0}$, the Gaussian (\ref{fizero}).
It then follows adapting the argument in \cite{Naza} that $\mathcal{H}%
_{\phi_{0}}\cap\mathcal{S}(Z)$ consists of all function $F$ such that 
\begin{equation}
\left( \tfrac{\partial}{\partial x_{j}}-i\tfrac{\partial}{\partial p_{j}}%
\right) (e^{\frac{1}{2}|z|^{2}}F(z))=0  \label{CR}
\end{equation}
\ for \ $1\leq j\leq n$. For arbitrary $\phi$ the space $\mathcal{H}_{\phi
}\cap\mathcal{S}(Z)$ is isometric to $\mathcal{H}_{\phi_{0}}\cap \mathcal{S}%
(Z)$.
\end{example}

\subsection{The operators $\widehat{A}_{\text{ph}}$}

Let us define operators $\widehat{T}_{\text{ph}}(z_{0})$ and $\widehat {A}_{%
\text{ph}}$ on $S^{\prime}(Z)$ by%
\begin{equation}
\widehat{T}_{\text{ph}}(z_{0})=e^{-\tfrac{i}{2}\sigma(\cdot,z_{0})}T(z_{0})
\label{hwnew}
\end{equation}
($T(z_{0})$ the translation operator in $Z$) and%
\begin{equation}
\widehat{A}_{\text{ph}}=\left( \tfrac{1}{2\pi}\right) ^{n}\int_{Z}a_{\sigma
}(z_{0})\widehat{T}_{\text{ph}}(z_{0})dz_{0}  \label{weyl2}
\end{equation}
with $a_{\sigma}=\mathcal{F}_{\sigma}a$.

\begin{example}
\label{oh}Let $a=$ $H$ be given by%
\begin{equation}
H=\frac{1}{2}(p^{2}+x^{2})\text{.}  \label{oha1}
\end{equation}
The corresponding operator is 
\begin{equation}
\widehat{H}_{\text{ph}}=-\frac{1}{2}\partial_{z}^{2}-i\frac{1}{2}%
\sigma(z,\partial_{z})+\frac{1}{8}|z|^{2}\text{.}  \label{oha2}
\end{equation}
\end{example}

Observe that the operators $\widehat{T}_{\text{ph}}$ satisfy the same
commutation relation as the usual Weyl--Heisenberg operators:%
\begin{equation}
\widehat{T}_{\text{ph}}(z_{1})\widehat{T}_{\text{ph}}(z_{0})=e^{-i\sigma
(z_{0},z_{1})}\widehat{T}_{\text{ph}}(z_{0})\widehat{T}_{\text{ph}}(z_{1})
\label{formuco2}
\end{equation}
and we have%
\begin{equation}
\widehat{T}_{\text{ph}}(z_{0})\widehat{T}_{\text{ph}}(z_{1})=e^{\frac{i}{2}%
\sigma(z_{0},z_{1})}\widehat{T}_{\text{ph}}(z_{0}+z_{1}).  \label{formuco3}
\end{equation}

Let $\mathbf{H}_{n}$ be the $(2n+1)$-dimensional Heisenberg group; it is the
set $Z\times\mathbb{R}$ equipped with the multiplicative law%
\begin{equation*}
(z,t)(z^{\prime},t^{\prime})=(z+z^{\prime},t+t^{\prime}+\tfrac{1}{2}%
\sigma(z,z^{\prime}))\text{.}
\end{equation*}
The \textquotedblleft Schr\"{o}dinger representation\textquotedblright\ of $%
\mathbf{H}_{n}$ is, by definition, the mapping $\widehat{T}$ which to every $%
(z_{0},t_{0})\in\mathbf{H}_{n}$ associates the unitary operator $\widehat {T}%
(z_{0},t_{0})$ on $L^{2}(X)$ defined by 
\begin{equation}
\widehat{T}(z_{0},t_{0})f(x)=\exp\left[ i(-t_{0}+\left\langle
p_{0},x\right\rangle -\tfrac{1}{2}\left\langle p_{0},x_{0}\right\rangle )%
\right] f(x-x_{0})\text{.}  \label{HG1}
\end{equation}
Recall that a classical theorem of Stone and von Neumann (see for instance
Wallach \cite{Wallach} for a modern detailed proof) says that the Schr\"{o}%
dinger representation is irreducible and that every irreducible unitary
representation of $\mathbf{H}_{n}$ is unitarily equivalent to $\widehat{T}$.
The relation (\ref{formuco3}) suggests that we define the phase-space
representation $\widehat{T}_{\text{ph}}$ of $\mathbf{H}_{n}$ in analogy with
(\ref{HG1}) by setting for $F\in L^{2}(Z)$ 
\begin{equation}
\widehat{T}_{\text{ph}}(z_{0},t_{0})F(z)=e^{it_{0}}\widehat{T}_{\text{ph}%
}(z_{0})F(z)\text{.}  \label{tps}
\end{equation}
Clearly $\widehat{T}_{\text{ph}}(z_{0},t_{0})$ is a unitary operator in $%
L^{2}(Z)$; moreover a straightforward calculation shows that%
\begin{equation}
\widehat{T}_{\text{ph}}(z_{0},t_{0})\widehat{T}_{\text{ph}}(z_{1},t_{1})=%
\widehat{T}_{\text{ph}}(z_{0}+z_{1},t_{0}+t_{1}+\tfrac{1}{2}%
\sigma(z_{0},z_{1}))  \label{formuco1}
\end{equation}
hence $\widehat{T}_{\text{ph}}$ is indeed a representation of the Heisenberg
group in $L^{2}(Z)$. We claim that:

\begin{proposition}
\label{thinter}(i) We have%
\begin{equation}
\widehat{T}_{\text{ph}}(z_{0},t_{0})U_{\phi}=U_{\phi}\widehat{T}(z_{0},t_{0})
\label{unit}
\end{equation}
hence the representation $\widehat{T}_{\text{ph}}$ is unitarily equivalent
to the Schr\"{o}dinger representation, and hence irreducible. (ii) The
following intertwining formula holds for every operator $\widehat{A}_{\text{%
ph}}$: 
\begin{equation}
\widehat{A}_{\text{ph}}U_{\phi}=U_{\phi}\widehat{A}.  \label{inter}
\end{equation}
\end{proposition}

\begin{proof}
Proof of \textit{(i)}. It suffices to prove that%
\begin{equation}
\widehat{T}_{\text{ph}}(z_{0})U_{\phi}=U_{\phi}\widehat{T}(z_{0})\text{.}
\label{intertwine}
\end{equation}
Let us write the operator $U_{\phi}$ in the form $U_{\phi}=e^{\frac{i}{2}%
\left\langle p,x\right\rangle }W_{\phi}$ that is%
\begin{equation}
W_{\phi}f(z)=\left( \tfrac{1}{2\pi}\right) ^{n/2}\int_{X}e^{-i\left\langle
p,x^{\prime}\right\rangle }\phi(x-x^{\prime})f(x^{\prime})dx^{\prime}\text{.}
\label{vefi}
\end{equation}
We have, by definition of $\widehat{T}_{\text{ph}}(z_{0})$ 
\begin{align*}
\widehat{T}_{\text{ph}}(z_{0})U_{\phi}f(z) & =\exp\left[ -\tfrac{i}{2}%
\sigma(z,z_{0})+\left\langle p-p_{0},x-x_{0}\right\rangle \right] W_{\phi
}f(z-z_{0}) \\
& =\exp\left[ \tfrac{i}{2}(-2\left\langle p,x_{0}\right\rangle +\left\langle
p_{0},x_{0}\right\rangle +\left\langle p,x\right\rangle )\right] W_{\phi
}f(z-z_{0})
\end{align*}
and, by definition of $W_{\phi}f$, 
\begin{align*}
W_{\phi}f(z-z_{0}) & =\left( \tfrac{1}{2\pi}\right) ^{n/2}\int
_{X}e^{-i\left\langle p-p_{0},x^{\prime}\right\rangle }\overline{\phi }%
(x-x^{\prime}-x_{0})f(x^{\prime})dx^{\prime} \\
& =\left( \tfrac{1}{2\pi}\right) ^{n/2}e^{i\left\langle
p-p_{0},x_{0}\right\rangle }\int_{X}e^{-i\left\langle
p-p_{0},x^{\prime\prime }\right\rangle }\overline{\phi}(x-x^{\prime%
\prime})f(x^{\prime\prime })dx^{\prime\prime}
\end{align*}
where we have set $x^{\prime\prime}=x^{\prime}+x_{0}$. The overall
exponential in $\widehat{T}_{\text{ph}}(z_{0})U_{\phi}f(z)$ is thus%
\begin{equation*}
u_{1}=\exp\left[ \frac{i}{2}(-\left\langle p_{0},x_{0}\right\rangle
+\left\langle p,x\right\rangle -2\left\langle
p,x^{\prime\prime}\right\rangle +2\left\langle
p_{0},x^{\prime\prime}\right\rangle )\right] \text{.}
\end{equation*}
Similarly,%
\begin{multline*}
U_{\phi}(\widehat{T}(z_{0})f)(z)=\left( \tfrac{1}{2\pi}\right) ^{n/2}e^{%
\frac{i}{2}\left\langle p,x\right\rangle }\times \\
\int_{X}e^{-i\left\langle p,x^{\prime\prime}\right\rangle }\overline{\phi }%
(x-x^{\prime\prime})e^{i(\left\langle p_{0},x^{\prime\prime}\right\rangle -%
\frac{1}{2}\left\langle p_{0},x_{0}\right\rangle
)}f(x^{\prime\prime}-x_{0})dx^{\prime\prime}
\end{multline*}
yielding the overall exponential%
\begin{equation*}
u_{2}=\exp\left[ i\left( \tfrac{1}{2}\left\langle p,x\right\rangle
-\left\langle p,x^{\prime\prime}\right\rangle +\left\langle
p_{0},x^{\prime\prime}\right\rangle -\tfrac{1}{2}\left\langle
p_{0},x_{0}\right\rangle \right) \right] =u_{1}
\end{equation*}
which proves (\ref{intertwine}). It follows from Stone--von Neumann's
theorem that $\widehat{T}_{\text{ph}}$ is an irreducible representation of $%
\mathbf{H}_{n}$ on each of the Hilbert spaces $\mathcal{H}_{\phi}$. Proof of 
\textit{(ii)}. In view of formula (\ref{intertwine}) we have%
\begin{align*}
\widehat{A}_{\text{ph}}U_{\phi}f & =\left( \tfrac{1}{2\pi}\right)
^{n}\int_{Z}a_{\sigma}(z_{0})\widehat{T}_{\text{ph}}(z_{0})U_{\phi}f(z)dz_{0}
\\
& =\left( \tfrac{1}{2\pi}\right) ^{n}\int_{Z}a_{\sigma}(z_{0})U_{\phi }(%
\widehat{T}(z_{0})f)(z)dz_{0} \\
& =\left( \tfrac{1}{2\pi}\right) ^{n}U_{\phi}\left( \int_{Z}a_{\sigma
}(z_{0})\widehat{T}(z_{0})f(z)dz_{0}\right) \\
& =U_{\phi}(\widehat{A}f)(z)
\end{align*}
hence (\ref{inter}).\bigskip
\end{proof}

Phase-space Weyl operators are composed in the usual way:

\begin{proposition}
Let $a_{\sigma}$ and $b_{\sigma}$ be the twisted symbols of the Weyl
operators $\widehat{A}_{\text{ph}}$ and $\widehat{B}_{\text{ph}}$. The
twisted symbol $c_{\sigma}$ of the compose $\widehat{A}_{\text{ph}}\widehat{B%
}_{\text{ph}}$ is the same as that of $\widehat{A}\widehat{B}$, that is%
\begin{equation*}
c_{\sigma}(z)=\left( \tfrac{1}{2\pi}\right) ^{n}\int e^{\frac{i}{2}%
\sigma(z,z^{\prime})}a_{\sigma}(z-z^{\prime})b_{\sigma}(z^{\prime})d^{2n}z%
\text{.}
\end{equation*}
\end{proposition}

\begin{proof}
By repeated use of (\ref{inter}) we have%
\begin{align*}
(\widehat{A}_{\text{ph}}\widehat{B}_{\text{ph}})U_{\phi} & =\widehat {A}_{%
\text{ph}}(\widehat{B}_{\text{ph}}U_{\phi}) \\
& =\widehat{A}_{\text{ph}}U_{\phi}\widehat{B} \\
& =U_{\phi}(\widehat{A}\widehat{B})
\end{align*}
hence $\widehat{A}_{\text{ph}}\widehat{B}_{\text{ph}}=(\widehat{A}\widehat {B%
})_{\text{ph}}$; the twisted symbol of $\widehat{A}\widehat{B}$ is precisely 
$c_{\sigma}$.\bigskip
\end{proof}

\subsection{Metaplectic covariance}

Let us now prove that the phase-space calculus enjoys a metaplectic
covariance property which is similar, \textit{mutandis mutatis}, to the
familiar corresponding property for usual Weyl operators (and which we will
discuss below); the latter is actually a straightforward consequence of the
intertwining relation%
\begin{equation}
\widehat{S}\widehat{T}(z_{0})\widehat{S}^{-1}=\widehat{T}(Sz_{0})
\label{interwh}
\end{equation}
valid for all $\widehat{S}\in\limfunc{Mp}(Z,\sigma)$and $z_{0}\in Z$.

We begin by noting that the restriction of the mapping $\widehat {A}%
\longrightarrow\widehat{A}_{\text{ph}}$ to $\limfunc{Mp}(Z,\sigma)$ is an
isomorphism of $\limfunc{Mp}(Z,\sigma)$ onto a subgroup $\limfunc{Mp}_{\text{%
ph}}(Z,\sigma)$ of the group of unitary operators on $L^{2}(Z)$. This
subgroup can thus be identified with the metaplectic group; the projection $%
\pi^{\limfunc{Mp}_{\text{ph}}}:\limfunc{Mp}_{\text{ph}}(Z,\sigma)%
\longrightarrow\limfunc{Mp}(Z,\sigma)$ is defined by 
\begin{equation*}
\pi^{\limfunc{Mp}_{\text{ph}}}(\widehat{S}_{\text{ph}})=\pi ^{\limfunc{Mp}}(%
\widehat{S})=S\text{.}
\end{equation*}

\begin{proposition}
Let $\widehat{S}_{\text{ph}}\in\limfunc{Mp}_{\text{ph}}(Z,\sigma)$ have
projection $S\in\limfunc{Sp}(Z,\sigma)$. Let $\widehat{A}$ have symbol $a$
and $\widehat{A_{S}}$ symbol $a\circ S$, $S\in\func{Sp}(Z,\sigma )$. We
have: 
\begin{equation}
\widehat{S}_{\text{ph}}\widehat{T}_{\text{ph}}(z_{0})\widehat{S}_{\text{ph}%
}^{-1}=\widehat{T}_{\text{ph}}(Sz)\text{ \ , \ }\widehat{A_{S}}_{\text{ph}}=%
\widehat{S}_{\text{ph}}^{-1}\widehat{A}_{\text{ph}}\widehat{S}_{\text{ph}}%
\text{.}  \label{metaco}
\end{equation}
\end{proposition}

\begin{proof}
Recall (formula (\ref{inter}) that $\widehat{A}_{\text{ph}}U_{\phi}=U_{\phi }%
\widehat{A}$; in particular we thus have $\widehat{S}_{\text{ph}}=U_{\phi }%
\widehat{S}U_{\phi}^{\ast}$ for every $\widehat{S}\in\limfunc{Mp}(Z,\sigma)$%
; it follows that%
\begin{equation*}
\widehat{S}_{\text{ph}}\widehat{T}_{\text{ph}}(z_{0})\widehat{S}_{\text{ph}%
}^{-1}=U_{\phi}\widehat{S}(U_{\phi}^{\ast}\widehat{T}_{\text{ph}%
}(z_{0})U_{\phi})\widehat{S}^{-1}U_{\phi}^{\ast}\text{.}
\end{equation*}
In view of formula (%
\index{intertwine}) we have%
\begin{equation*}
U_{\phi}^{\ast}%
\widehat{T}_{\text{ph}}(z_{0})U_{\phi}=\widehat{T}(z_{0})
\end{equation*}
and hence, by (\ref{interwh}),%
\begin{align*}
\widehat{S}_{\text{ph}}\widehat{T}_{\text{ph}}(z_{0})\widehat{S}_{\text{ph}%
}^{-1} & =U_{\phi}\widehat{S}\widehat{T}(z_{0})\widehat{S}^{-1}U_{\phi
}^{\ast} \\
& =U_{\phi}\widehat{T}(Sz_{0})U_{\phi}^{\ast} \\
& =\widehat{T}_{\text{ph}}(Sz)
\end{align*}
which proves the first formula (\ref{metaco}). The second formula is proven
in the same way using the equalities $\widehat{A}_{\text{ph}}=U_{\phi}%
\widehat {A}U_{\phi}^{\ast}$: we have%
\begin{align*}
\widehat{S}_{\text{ph}}^{-1}\widehat{A}_{\text{ph}}\widehat{S}_{\text{ph}} &
=(\widehat{S}_{\text{ph}}^{-1}U_{\phi})\widehat{A}(U_{\phi}^{\ast}\widehat {S%
}_{\text{ph}}) \\
& =U_{\phi}(\widehat{S}^{-1}\widehat{A}\widehat{S})U_{\phi}^{\ast} \\
& =U_{\phi}^{\ast}\widehat{A_{S}}U_{\phi}^{\ast}
\end{align*}
hence the result since $U_{\phi}^{\ast}\widehat{A_{S}}U_{\phi}^{\ast}=%
\widehat{A_{S}}_{\text{ph}}$. (Alternatively we could have proven the second
formula (\ref{metaco}) using the first together with definition (\ref{weyl2}%
) of $\widehat{A}_{\text{ph}}$.\bigskip
\end{proof}

Let us shortly discuss the meaning of this result for the uniqueness of the
phase-space Weyl calculus we have constructed in this paper.

In \cite{Shale} Shale proves the following result (see \cite{Wong}, Chapter
30 for a detailed proof): let 
\begin{equation*}
\mathcal{L}_{X}=\mathcal{L}(\mathcal{S}(X),\mathcal{S}^{\prime}(X))
\end{equation*}
be the set of all continuous linear mappings $\mathcal{S}(X)\longrightarrow 
\mathcal{S}^{\prime}(X)$. Let $\limfunc{Op}:\mathcal{S}^{\prime
}(Z)\longrightarrow\mathcal{L}_{X}$ be a sequentially continuous mapping
such that:

\begin{itemize}
\item We have%
\begin{equation}
\limfunc{Op}(a)f(x)=a(x)f(x)  \label{trivial}
\end{equation}
if $f\in\mathcal{S}(X)$ and $a\in L^{\infty}(X)\subset\mathcal{S}%
^{\prime}(Z) $;

\item We have 
\begin{equation}
\widehat{S}\limfunc{Op}(a)\widehat{S}^{-1}=\limfunc{Op}(a\circ S^{-1})
\label{soppa}
\end{equation}
for every $\widehat{S}\in\limfunc{Mp}(Z,\sigma)$ with $S=\pi ^{\limfunc{Mp}}(%
\widehat{S})$.
\end{itemize}

Then $\limfunc{Op}(a)=\widehat{A}$, the Weyl operator associated with $a$.
In other words, the metaplectic covariance property (\ref{soppa}) uniquely
characterizes the class of operators $\mathcal{S}(X)\longrightarrow \mathcal{%
S}^{\prime}(X)$ which in addition satisfies the triviality condition (\ref%
{trivial}).

A straightforward duplication of Shale's proof leads to the following
statement:

\begin{proposition}
Let $\mathcal{L}_{Z}=\mathcal{L}(\mathcal{S}(Z),\mathcal{S}^{\prime}(Z))$ be
the set of all continuous linear mappings $\mathcal{S}(Z)\longrightarrow 
\mathcal{S}^{\prime}(Z)$. Let $\limfunc{Op}_{\text{ph}}:\mathcal{S}%
^{\prime}(Z)\longrightarrow\mathcal{L}_{Z}$ be a sequentially continuous
mapping such that $\limfunc{Op}_{\text{ph}}(1)$ is the identity and%
\begin{equation*}
\widehat{S}_{\text{ph}}\limfunc{Op}\nolimits_{\text{ph}}(a)\widehat {S}_{%
\text{ph}}^{-1}=\limfunc{Op}\nolimits_{\text{ph}}(a\circ S^{-1})\text{.}
\end{equation*}
Then $\limfunc{Op}_{\text{ph}}(a)=\widehat{A}_{\text{ph}}$.
\end{proposition}

That $\widehat{A}_{\text{ph}}=I$ if $a=1$ immediately follows from the
observation that $F_{\sigma}1=(2\pi)^{n}\delta$ where $\delta$ is the Dirac
distribution on $Z$ so that%
\begin{equation*}
\widehat{A}_{\text{ph}}F(z)=\int_{Z}\delta(z_{0})\widehat{T}_{\text{ph}%
}(z_{0})F(z)dz=F(z)\text{.}
\end{equation*}

\section{Conclusions and Perspectives}

Let us begin with the perspectives. The Weyl--Wigner--Moyal formalism (and
in particular Weyl calculus in its modern form) originates in the efforts of
generations of mathematicians (and physicists) to provide quantum mechanics
with an efficient and rigorous framework to \textquotedblleft
quantize\textquotedblright\ functions into operators (or, on a subtler and
more useful level, to \textquotedblleft dequantize\textquotedblright\
operators, see \cite{Mackey}). What could be the advantages (or
disadvantages) of using the phase-space calculus introduced in this article?
I have mainly in mind the applications to quantum mechanics; while it is
difficult to argue that there are practical advantages in solving the
phase-space Schr\"{o}dinger equation 
\begin{equation}
i\partial_{t}\Psi(z)=\widehat{H}_{\text{ph}}\Psi(z)  \label{sch1}
\end{equation}
instead of the usual%
\begin{equation}
i\partial_{t}\psi(x)=\widehat{H}\psi(x)  \label{sch2}
\end{equation}
(the first depends on $2n+1$ variables and the second on only $n+1$
variable) there are interesting conceptual issues that arise. While it is
clear that the solutions of (\ref{sch2}) are taken into solutions of (\ref%
{sch1}) using any of the isometries $U_{\phi}:L^{2}(X)\longrightarrow
L^{2}(Z)$, the converse is not true. We have discussed in \cite{jphysa}
(somewhat in embryonic form) the interpretation of general solutions of (\ref%
{sch1}); since there is no point in duplicating these results we refer the
interested reader to that paper. Suffice it to say that under sufficient
assumptions on their support the Gaussian functions $\Psi\in L^{2}(Z)$ can
be viewed as the Wigner transforms of general Gaussian \textquotedblleft
mixed states\textquotedblright. In the general case there is still much work
to do; we hope to come back to this topic in a near future.

Let us finally indicate a few connections between our approach and results
from other authors.

In \cite{Howe} Howe defines and studies the \textquotedblleft oscillator
semi-group\textquotedblright\ $\Omega $. It is the semi-group of Weyl
operators whose symbols are arbitrary centered Gaussians; we we have only
considered \ symbols which are Gaussians having purely imaginary exponents.
One of the main differences between our approach and Howe's lies in the
treatment of the metaplectic representation: in \cite{Howe} its study is
performed by moving to Fock space, which allows Howe to bypass the
difficulties occurring when $\widehat{S}\in \limfunc{Mp}(Z,\sigma )$ is no
longer of the type $\widehat{S}_{W,m}$ (see the comments in \cite{Folland},
p. 246). In the present work these difficulties are solved in a more
explicit way by writing $\widehat{S}$ as a product $\widehat{S}_{W,m}%
\widehat{S}_{W^{\prime },m^{\prime }}$ with $\det (S_{W}-I)\neq 0$, $\det
(S_{W^{\prime }}-I)\neq 0$ (Corollary \ref{propabove}); this allows us to
determine explicitly the correct phase factor $i^{\nu }$ in the Weyl
representation of $\widehat{S}$ (which is closely related to the
Conley--Zehnder index) by using the powerful machinery of the ALM\ index.
(Let us mention, in passing, that the factorization $\widehat{S}=\widehat{S}%
_{W,m}\widehat{S}_{W^{\prime },m^{\prime }}$, which goes back to Leray \cite%
{Leray}, does not seem to be widely known by mathematicians working on the
metaplectic representation; it can however easily be proven noting that the
symplectic group acts transitively on pairs of transverse Lagrangian planes;
see \cite{Wiley}).

An early version of the operators $\widehat{R}_{\nu}(S)$ has appeared in the
work of Mehlig and Wilkinson \cite{WM}; it was this paper which triggered
the present author's interest in the study of the Weyl symbol of metaplectic
operators; for an early version see \cite{mdglettmath}. Mehlig and
Wilkinson's primary goal is to establish trace formulae related to the
Gutzwiller approach to semi-classical quantum systems \cite{CRL}: the
precise determination of the Conley--Zehnder-type index $\nu$ could
certainly be of some use in such a project (but the roadblocks on the way to
a rigorous and complete theory are still immense, in spite of many attempts
and some advances, see for instance \cite{CRL}).

The choice we did not make for reasons explained in the beginning of the
Introduction --namely the use of the standard Heisenberg--Weyl operators $%
\widehat{T}(z_{0})$ extended to phase space-- leads on the
quantum-mechanical level to the Schr\"{o}dinger equation written formally as%
\begin{equation*}
i\partial_{t}\Psi(z)=H(x+i\partial_{p},-i\partial_{x})\Psi(z)\text{;}
\end{equation*}
the latter has been obtained using non-rigorous \textquotedblleft coherent
state representation\textquotedblright\ arguments by Torres-Vega and
Frederick \cite{TV}, and is currently being an object of lively discussions
in physics circles; see our comments and references in \cite{extend}.

It would perhaps be interesting to recast some of our results in the more
general setting of abstract harmonic analysis and representation theory
considered in \cite{bohan,Hannabuss}, where formal similarities with the
present work are to be found (I take the opportunity to thank K. Hannabuss
for having drawn my attention to his work on the topic). The
\textquotedblleft quantization rules\textquotedblright\ (\ref{quantumrule})
also have a definite resemblance with formulae appearing in deformation
quantization \textit{\`{a} la }Bayen \textit{et al.} \cite{BFFLS}; since the
latter is (in its simplest case) based on the notion of Moyal star-product,
itself related to the Wigner--Moyal--Weyl formalism, this is not \textit{a
priori} surprising: it is very possible that both approaches are cousins,
even if obtained by different methods.\bigskip\


\begin{thebibliography}{99}
\bibitem{BFFLS} F. Bayen, M. Flato, C. Fronsdal, A.\ Lichnerowicz, and D.\
Sternheimer. Deformation Theory and Quantization. I. Deformation of
Symplectic Structures. Annals of Physics 111 (1978) 6--110; II Physical
Applications 110 (1978) 111--151.

\bibitem{BF1} B. Booss--Bavnbek and K. Furutani. The Maslov Index: a
Functional Analytical Definition and the Spectral Flow Formula, Tokyo J.
Math. 21(1) (1998).

\bibitem{bohan} D.\ Bowes and K. Hannabuss. Weyl quantization and star
products, J. Geom. Phys.\ 22\textbf{\ }(1997), 319--348.

\bibitem{CLM} S. E. Cappell, R. Lee and E. Y. Miller. On the Maslov index,
Comm. Pure and Appl. Math\textit{.} 17\textbf{\ (}1994), 121--186.

\bibitem{Ciriza} E. Ciriza. Bifurcation of periodic orbits of time dependent
Hamiltonian systems on symplectic manifolds, Rend. Sem. Mat. Univ. Pol.
Torino 57(3) (1999), 161--173.

\bibitem{CZ} C. Conley and E. Zehnder. Morse-type index theory for flows and
periodic solutions of Hamiltonian equations. Comm. Pure and Appl. Math. 37
(1984) 207--253.

\bibitem{CRL} S. C. Creagh, J. M. Robbins, and R.\ G. Littlejohn.
Geometrical properties of Maslov indices in the semiclassical trace formula
for the density of states. Phys. Rev. A 42(4) (1990) 1907--1922.

\bibitem{Dazord} P. Dazord. Invariants homotopiques attach\'{e}s aux fibr%
\'{e}s symplectiques. Ann. Inst.\textit{\ }Fourier. 29(2) (1979) 25--78.

\bibitem{Folland} G. B. Folland. Harmonic Analysis in Phase space. Annals of
Mathematics studies, Princeton University Press, Princeton, N.J. 1989.

\bibitem{GPP} R. Giambo, P. Piccione, and A. Portaluri. Computation of the
Maslov index and the spectral flow via partial signatures. C. R. Math. Acad.
Sci. Paris. 338(5) (2004), 397--402; Partial signatures, spectral flow and
the Maslov index. Semi-Riemannian index theorems in the degenerate case. To
appear in Comm. Analysis and Geometry.

\bibitem{AIF} M. de Gosson. Maslov Indices on $\limfunc{Mp}(n)$.\textit{\ }%
Ann. Inst. Fourier 40(3) (1990) 537--55.

\bibitem{JMPA} M. de Gosson. The structure of $q$-symplectic geometry. J.
Math. Pures et Appl. 71 (1992), 429--453.

\bibitem{Wiley} M. de Gosson. Maslov Classes\textit{, }Metaplectic
Representation and Lagrangian\textit{\ }Quantization. Research Notes in
Mathematics. 95, Wiley--VCH, Berlin, 1997.

\bibitem{AMS} M. de Gosson. Lagrangian path intersections and the Leray
index. In Contemp.\textit{\ }Math. 258, 2000.

\bibitem{MSDG1} M. de Gosson and S. de Gosson. Symplectic path intersections
and the Leray index. Progr. Nonlinear Differential Equations Appl. 52, Birkh%
\"{a}user. 2003.

\bibitem{MSDG2} M. de Gosson and S. de Gosson. The cohomological
interpretation of the indices of Robbin and Salamon. Jean Leray 
%TCIMACRO{\U{b4}}%
%BeginExpansion
\'{}%
%EndExpansion
99 Conference Proceedings, Math. Phys. Studies 4, Kluwer Academic Press.
2003.

\bibitem{mdglettmath} M. de Gosson. The Weyl Representation of Metaplectic
operators. Letters in\textit{\ }Mathematical Physics 72 (2005) 129--142.

\bibitem{bullscimath} M. de Gosson. Cellules quantiques symplectiques et
fonctions de Husimi--Wigner. Bull. Sci. Math. 129 (2005) 211--226.

\bibitem{extend} M. de Gosson. Extended Weyl Calculus and Application to the
Phase-Space Schr\"{o}dinger Equation. J. Phys. A: Math. and General 38
(2005) 325--329.

\bibitem{jphysa} M. de Gosson. Symplectically Covariant Schr\"{o}dinger
Equation in Phase Space. J. Phys. A. Math. and Gen. (2005) (in press).

\bibitem{Gro} K. Gr\"{o}chenig. Foundations of Time-Frequency Analysis. Birkh%
\"{a}user, Boston. 2000.

\bibitem{Grossmann} A. Grossmann, G. Loupias, and E. M. Stein. An algebra of
pseudo-differential\textit{\ }operators and quantum mechanics in phase
space, Ann. Inst. Fourier. Grenoble 18(2) (1968) 343--368.

\bibitem{Hannabuss} K. C. Hannabuss. Characters and contact transformations.
Math. Proc. Camb. Phil.\textit{\ }Soc. 90 (1981) 465--476.

\bibitem{HWZ} H. Hofer, K. Wysocki, and E. Zehnder. Properties of
pseudoholomorphic curves in symplectizations I: Asymptotics. Ann. Inst.
Henri Poincar\'{e}. Analyse non lin\'{e}aire. 13(3) (1996) 337--379; II:
Embedding controls and algebraic invariants. Geometric and Functional
Analysis 2(5) (1995) 270--328.

\bibitem{HZ} H. Hofer and E. Zehnder. Symplectic Invariants and Hamiltonian
Dynamics. Birkh\"{a}user Advanced texts Basler Lehrb\"{u}cher, Birkh\"{a}%
user Verlag. 1994.

\bibitem{Howe} R. Howe. The Oscillator semigroup. Proc. of Symposia in Pure
Mathematics 48\textbf{\ }61--132, Amer. Math. Soc. 1988.

\bibitem{Leray} J. Leray. Lagrangian Analysis and Quantum Mechanics,\ a
mathematical structure\textit{\ }related\textit{\ }to asymptotic expansions
and the Maslov index. MIT Press, Cambridge, Mass. (1981); translated from:
Analyse Lagrangienne RCP 25, Strasbourg and Coll\`{e}ge de France.
1976--1977.

\bibitem{LV} G. Lion, and M. Vergne. The Weil representation, Maslov index
and Theta series. Progress in mathematics 6, Birkh\"{a}user. 1980.

\bibitem{Littlejohn} R. G. Littlejohn. The semiclassical evolution of wave
packets. Physics Reports 138\textbf{(}4--5) (1986), 193--291.

\bibitem{Duff} D. McDuff and D. Salamon. Symplectic Topology.\textit{\ }%
Oxford Science Publications, 1998.

\bibitem{WM} B. Mehlig and M. Wilkinson. Semiclassical trace formulae using
coherent states. Ann. Phys. 18(10), 6--7 (2001) 541--555.

\bibitem{Mackey} G. W. Mackey. The Relationship Between Classical and
Quantum Mechanics. In Contemporary Mathematics 214, Amer. Math. Soc.
Providence, RI, 1998.

\bibitem{Morse} M. Morse. The Calculus of Variations in the Large. AMS,
Providence, R. I, 1935.

\bibitem{Naza} V Nazaikiinskii, B W Schulze, and B Sternin, Quantization
Methods in Differential Equations (Taylor \& Francis, 2002).

\bibitem{Piccione2} R. C. Nostre Marques, P. Piccione, and D. Tausk. On the
Morse and the Maslov index for periodic geodesics of arbitrary causal
character. Differential Geometry and its Applications. Proc. Conf. Opava,
2001.

\bibitem{SZ} D. A. Salamon and E. Zehnder. Morse theory for periodic
solutions of Hamiltonian systems and the Maslov index. Comm. Pure and Appl.
Math. 45\textbf{\ (}1992) 1303--1360.

\bibitem{RS} J. Robbin and D. Salamon. The Maslov index for paths. Topology
32 (1993) 827--844.

\bibitem{Shale} D. Shale. Linear Symmetries of free Boson fields. Trans.
Amer. Math. Soc. 103 (1962) 149--167.

\bibitem{Souriau} J.-M. Souriau. Construction explicite de l'indice de
Maslov. In: Group Theoretical\textit{\ }Methods in Physics, Lecture Notes in
Physics, 50 (1975) 17--148, Springer-Verlag.

\bibitem{Stein} E. M. Stein. Harmonic Analysis: Real Variable Methods,
Orthogonality, and Oscillatory\textit{\ }Integrals. Princeton University
Press. 1993.

\bibitem{TV} G. Torres-Vega and J. H. Frederick. Quantum mechanics in phase
space: New approaches to the correspondence principle. J. Chem. Phys. 93(12)
(1990) 8862--8874.

\bibitem{Wall} C. T. C. Wall. Nonadditivity of the signature. Invent. Math.
7, 1969, 269--274.

\bibitem{Wallach} Wallach, N. Lie Groups: History, Frontiers and
Applications, 5. Symplectic Geometry and Fourier Analysis, Math Sci Press,
Brookline, MA. 1977.

\bibitem{Wong} M. W. Wong. Weyl Transforms\textit{.} Springer. 1998.
\end{thebibliography}
\end{document}